%% file: 0_CHI2026.tex
\documentclass[manuscript]{acmart}
\AtBeginDocument{%
  }

\usepackage{csquotes}
\usepackage{enumitem}
\usepackage{array}
\usepackage{natbib}
\usepackage{booktabs,tabularx}
\usepackage{xcolor}
\usepackage{framed}
\newcolumntype{P}[1]{>{\centering\arraybackslash}p{#1}}

\newcommand\revision[1]{%
   {\sf\textcolor{blue}{#1}}%
}

\renewcommand{\revision}[1]{#1}

\newcommand{\nod}{\textbf{\texttt{No Disclosure}}}    
\newcommand{\partd}{\textbf{\texttt{Partial Disclosure}}}
\newcommand{\fulld}{\textbf{\texttt{Full Disclosure}}}

\newcommand{\lowden}{\textbf{\texttt{Low Density}}}    
\newcommand{\medden}{\textbf{\texttt{Medium Density}}}
\newcommand{\highden}{\textbf{\texttt{High Density}}}

\copyrightyear{2026}
\acmYear{2026}
\setcopyright{cc}
\setcctype{by}
\acmConference[CHI '26]{Proceedings of the 2026 CHI Conference on Human Factors in Computing Systems}{April 13--17, 2026}{Barcelona, Spain}
\acmBooktitle{Proceedings of the 2026 CHI Conference on Human Factors in Computing Systems (CHI '26), April 13--17, 2026, Barcelona, Spain}
\acmPrice{}
\acmDOI{10.1145/3772318.3791420}
\acmISBN{979-8-4007-2278-3/2026/04}

\begin{document}

\title[How Information Asymmetries About AI System Capabilities Affect Market Outcomes and Adoption]{When Life Gives You AI, Will You Turn It Into A Market for Lemons? 
Understanding How Information Asymmetries About AI System Capabilities Affect Market Outcomes and Adoption
}

\author{Alexander Erlei}
\email{alexander.erlei@wiwi.uni-goettingen.de}
\orcid{0000-0001-7322-2761}
\affiliation{%
  \institution{University of Göttingen}
  \city{Göttingen}
  \country{Germany}
}

\author{Federico Cau}
\email{federicom.cau@unica.it}
\orcid{0000-0002-8261-3200}
\affiliation{%
  \institution{University of Cagliari}
  \city{Cagliari}
  \country{Italy}
}

\author{Radoslav Georgiev}
\email{r.k.georgiev@student.tudelft.nl}
\orcid{0009-0009-2649-8621}
\affiliation{%
  \institution{Delft University of Technology}
  \city{Delft}
  \country{The Netherlands}
}

\author{Sagar Kumar}
\email{sagar.chethankumar@columbia.edu}
\orcid{0009-0008-7006-5148}
\affiliation{%
  \institution{Columbia University}
  \city{New York}
  \country{USA}
}

\author{Kilian Bizer}
\email{bizer@wiwi.uni-goettingen.de}
\orcid{0000-0002-2258-7725}
\affiliation{%
  \institution{University of Göttingen}
  \city{Göttingen}
  \country{Germany}
}

\author{Ujwal Gadiraju}
\email{u.k.gadiraju@tudelft.nl}
\orcid{0000-0002-6189-6539}
\affiliation{%
  \institution{Delft University of Technology}
  \city{Delft}
  \country{The Netherlands}
}


\begin{abstract}
  AI consumer markets are characterized by severe buyer-supplier market asymmetries. Complex \revision{AI} systems can appear highly accurate while making costly errors or embedding hidden defects. While there have been regulatory efforts surrounding different forms of disclosure, large information gaps remain. \revision{This paper provides the first experimental evidence on the important role of information asymmetries and disclosure designs in shaping user adoption of AI systems.} We systematically vary the density of low-quality AI systems and the depth of disclosure requirements in a simulated AI product market to gauge how people react to the risk of accidentally \revision{relying on a low-quality AI system}. Then, we compare participants’ choices to a rational Bayesian model, analyzing the degree to which partial information disclosure can improve AI adoption. Our results underscore the deleterious effects of information asymmetries on AI adoption, but also highlight the potential of \revision{partial disclosure designs to improve the overall efficiency of human decision-making.} 
\end{abstract}



\begin{teaserfigure}\includegraphics[width=\textwidth]{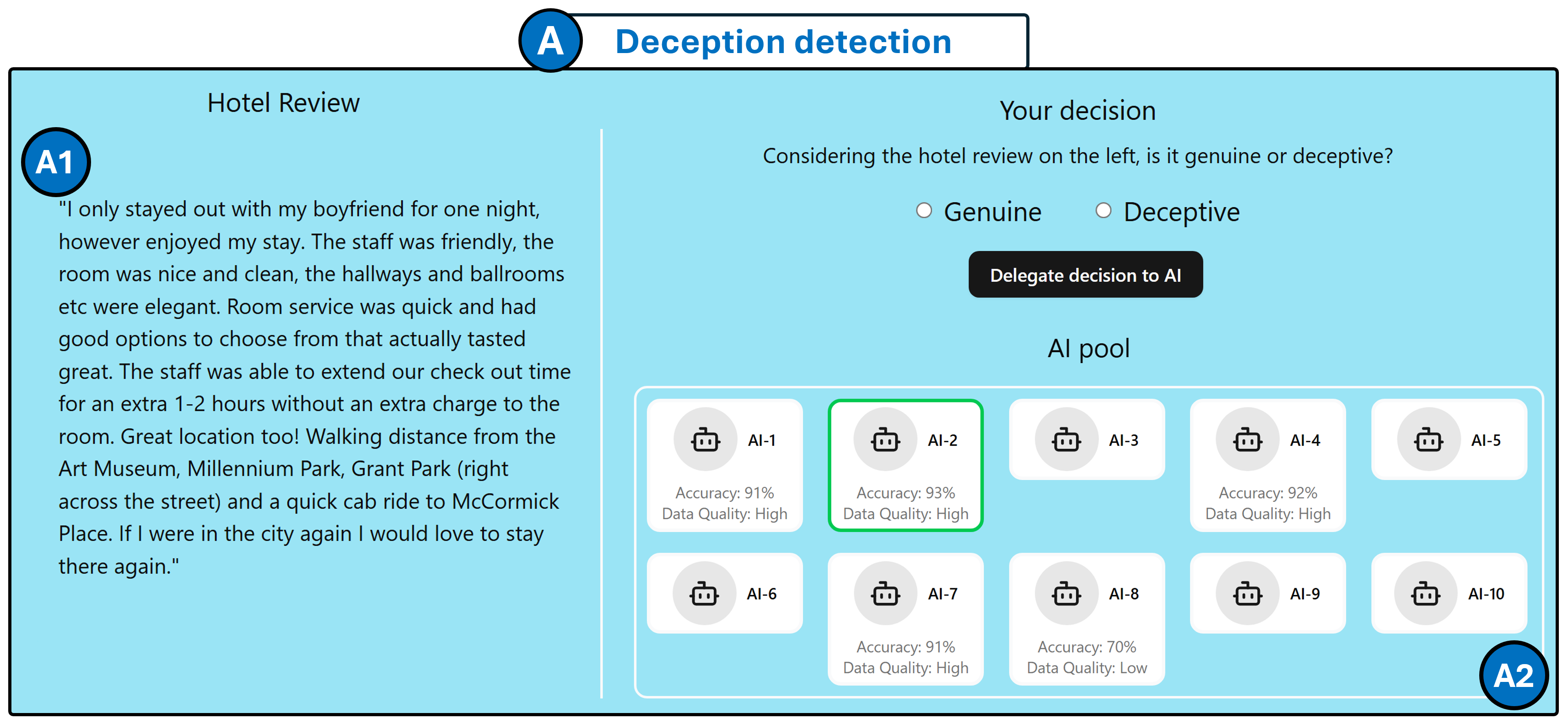}
  \caption{Task interface that participants interacted with during the study. This illustration represents the \textit{deceptive review detection} task (A) with full information disclosure about the AI systems in the available pool. On the left side (A1), participants viewed the hotel review they needed to verify. 
  On the right side, participants saw the binary decision options for the current trial, and an AI pool of ten AI systems (A2). By hovering over an AI system, participants could access  
  the available information (\textit{i.e.,} quality indicators) corresponding to the AI system in conditions with partial or full information disclosure. Participants could select an AI system from the pool for delegation by first clicking on it---which makes the selected AI system light up with a green border---and then clicking the `\textit{Delegate decision to AI}' button to complete the current trial.}
  \Description{Figure 1 depicts the interface used by participants for the deceptive review detection task. The screen is divided into two main sections. On the left, participants read the hotel review they need to evaluate. On the right, they see two binary response options (i.e., classifying the review as either genuine or deceptive) and a pool of ten AI systems represented as individual tiles. When participants hover over a tile, the system displays the available quality indicators for that AI model with partial (i.e., AI accuracy) or full (i.e., AI accuracy and data quality) information disclosure. Participants can either make the decision themselves or delegate it by selecting one of the AI systems, which becomes highlighted with a green border, and then clicking the \enquote{Delegate decision to AI} button.
}
  \label{fig:teaser}
\end{teaserfigure}


\maketitle

\input{1_New_Introduction}
\input{2_RelatedWork}

\input{3_Method}
\input{4_Results}

\input{5_Discussion}

\input{6_Conclusions}

\begin{acks}
We thank all the anonymous participants in our study. This work was partially supported by the TU Delft AI Initiative, the Model Driven Decisions Lab (\textit{MoDDL}), the Robust LTP GENIUS Lab, and the \textit{ProtectMe} Convergence Flagship. 
\end{acks}

\bibliographystyle{ACM-Reference-Format}
\bibliography{references}

\clearpage

\section*{Appendix}

\begin{table*}[ht]
\centering
\caption{Summary of participants'  task familiarity across conditions. Mean (M) and standard deviation (SD) are reported for each of the three tasks (loan approval, deceptive review detection, and skin cancer prediction). $Sample$ indicates the number of observations per condition.}
\small
\setlength{\tabcolsep}{6pt} 
\begin{tabularx}{\textwidth}{
>{\raggedright\arraybackslash}X
>{\centering\arraybackslash}p{1.0cm} >{\centering\arraybackslash}p{0.1cm} 
>{\centering\arraybackslash}p{0.9cm} >{\centering\arraybackslash}p{0.9cm} 
>{\centering\arraybackslash}p{0.9cm} >{\centering\arraybackslash}p{0.9cm} 
>{\centering\arraybackslash}p{0.9cm} >{\centering\arraybackslash}p{0.9cm} 
}
\toprule
Condition & Sample & & Loan$_{\textnormal{M}}$ & Loan$_{\textnormal{SD}}$ & Reviews$_{\textnormal{M}}$ & Reviews$_{\textnormal{SD}}$ & Cancer$_{\textnormal{M}}$ & Cancer$_{\textnormal{SD}}$ \\
\midrule
Full disclosure, high density      & 30 & & 2.43 & 1.36 & 2.03 & 1.00 & 1.60 & 0.89 \\
No disclosure, low density        & 50 & & 2.40 & 1.41 & 2.10 & 1.20 & 1.78 & 1.13 \\
No disclosure, medium density     & 50 & & 2.32 & 1.19 & 1.90 & 1.05 & 1.50 & 0.74 \\
No disclosure, high density       & 50 & & 2.42 & 1.18 & 2.00 & 1.21 & 1.60 & 0.97 \\
Partial disclosure, low density   & 50 & & 2.46 & 1.18 & 2.12 & 1.21 & 1.76 & 0.87 \\
Partial disclosure, medium density& 50 & & 2.36 & 1.17 & 2.14 & 1.16 & 1.56 & 0.93 \\
Partial disclosure, high density  & 50 & & 2.06 & 1.11 & 1.86 & 0.97 & 1.58 & 0.99 \\
\bottomrule
\end{tabularx}
\label{tab:task_familiarity}

\vspace{0.5em}
\footnotesize
\emph{Notes.} We assessed the distribution of task familiarity scores for loan approval, deceptive reviews, and skin cancer tasks using the Shapiro–Wilk normality test. All three tasks significantly deviated from normality (loan approval: $W = 0.871$, $p < .001$; deceptive reviews: $W = 0.819$, $p < .001$; skin cancer: $W = 0.699$, $p < .001$). Consequently, we applied non-parametric Kruskal–Wallis tests to examine potential differences across conditions. Results indicated no significant differences in task familiarity across conditions for any of the tasks (loan approval: $\chi^2 = 3.64$, $df = 6$, $p = 0.72$; deceptive reviews: $\chi^2 = 2.30$, $df = 6$, $p = 0.89$; skin cancer: $\chi^2 = 3.89$, $df = 6$, $p = 0.69$).

\end{table*}

\begin{figure}[h]
    \centering
    \includegraphics[width=0.45\textwidth]{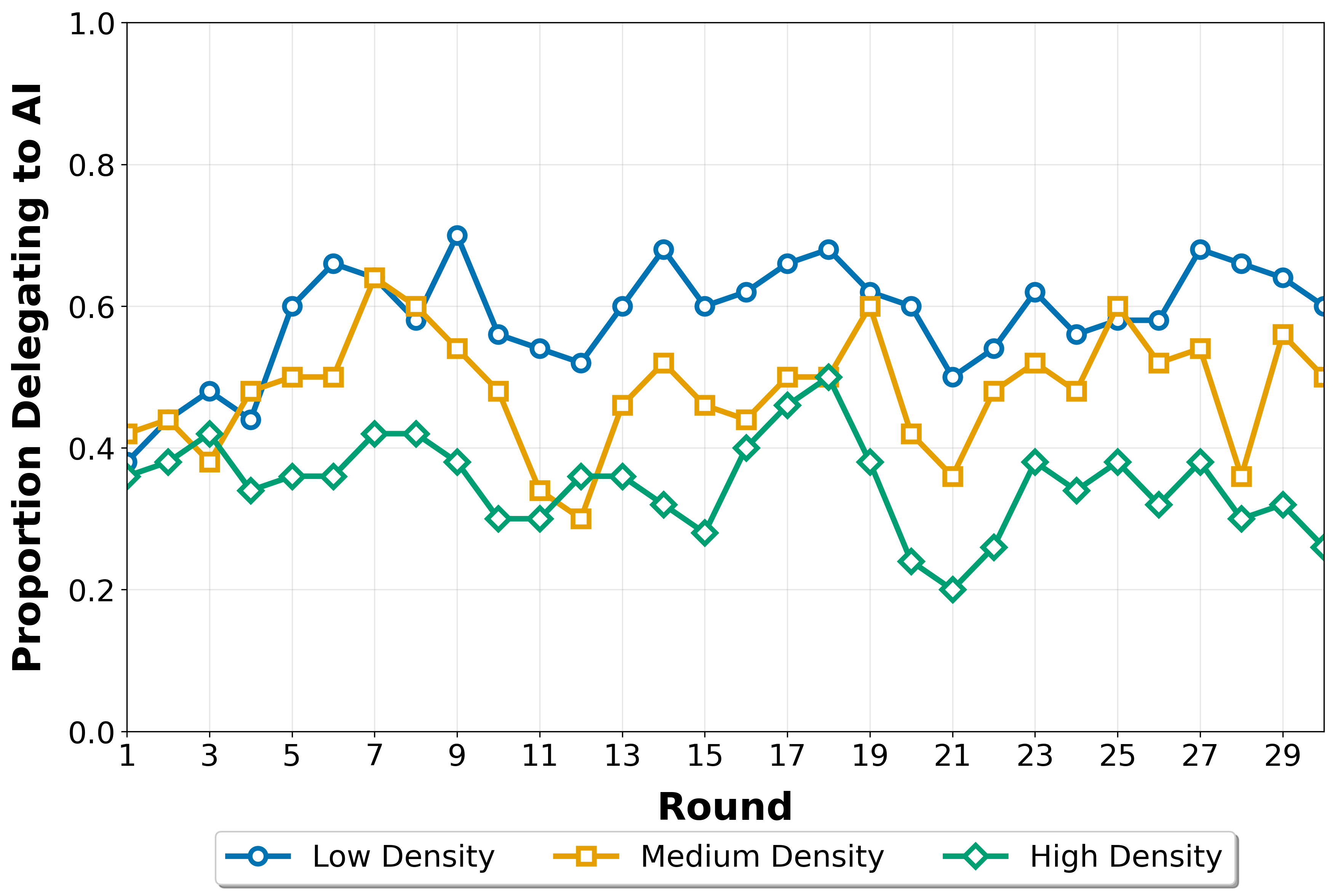}
    \caption{Average subject delegation in \nod{} between density conditions.}
    \label{fig:a1}
\end{figure}

\begin{figure}[h]
    \centering
    \includegraphics[width=0.45\textwidth]{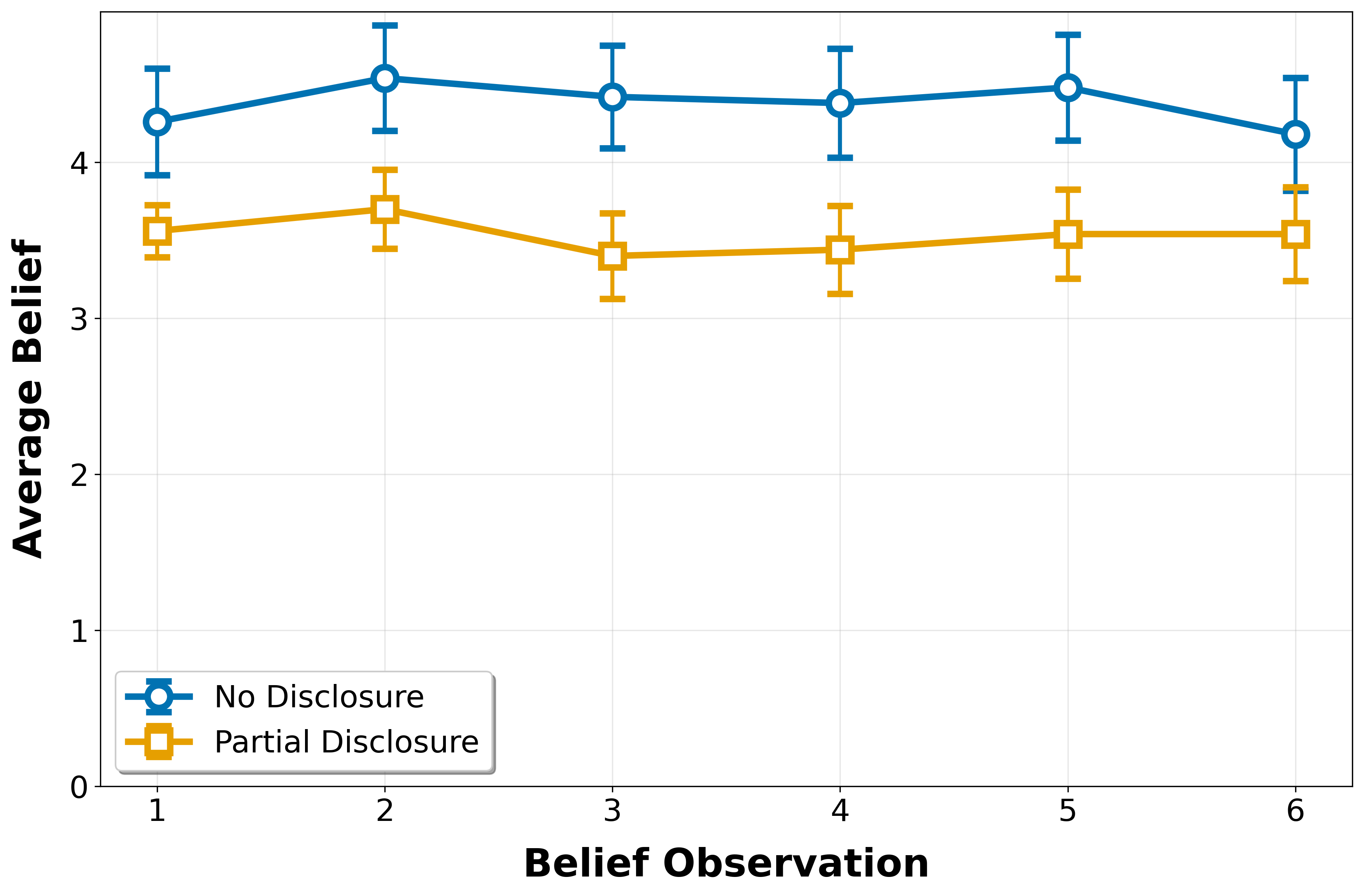}
    \includegraphics[width=0.45\textwidth]{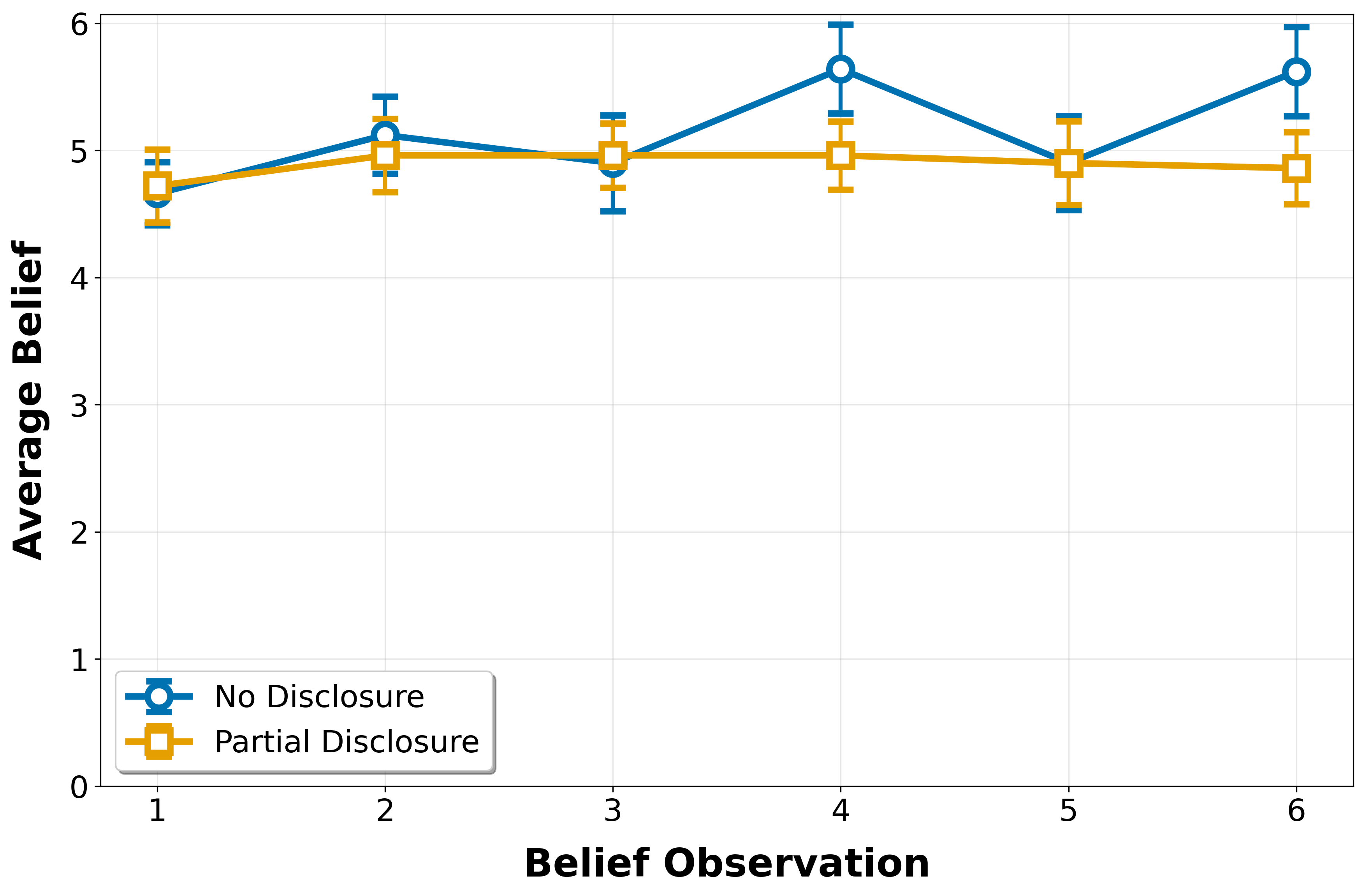}
    \includegraphics[width=0.45\textwidth]{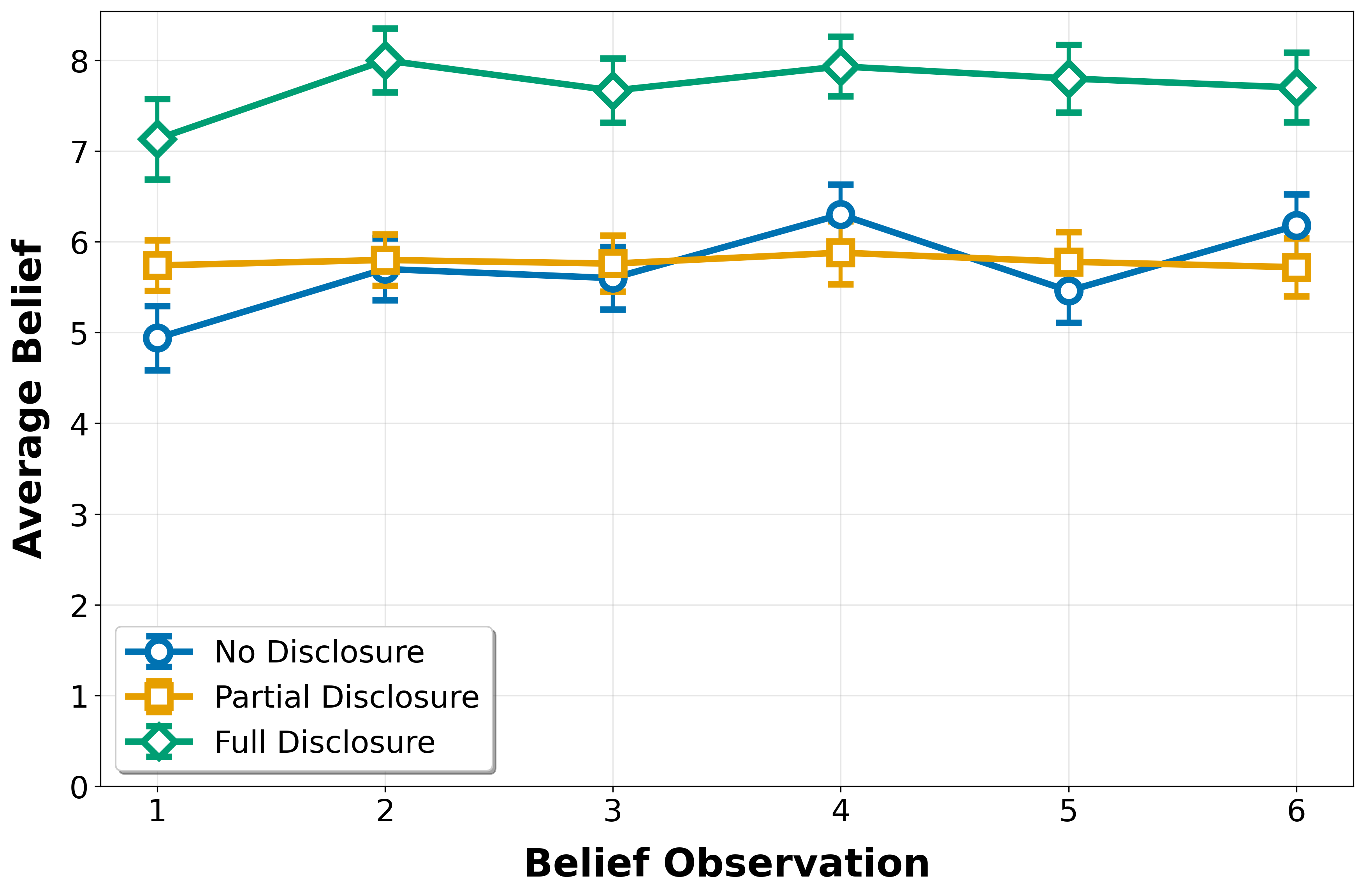}
    
    \caption{Belief evolution over time in order \lowden{}, \medden{}, \highden{}.}
    \label{fig:beliefs}
\end{figure}

\begin{table}[t]
\centering
\caption{Time trend in delegation under \nod{}: mixed-effects logit with subject RE, task FE, and density$\times$round interaction.}
\label{tab:a_01}
\begin{tabular}[h]{l c}
\toprule
 & \textbf{Delegation Share (Log Odds)} \\
\midrule
\textbf{Density (vs Low)} \\
\quad Medium & $-0.277\,(0.359)$ \\
\quad High   & $-0.743^{*}\,(0.359)$ \\
\addlinespace[3pt]
\textbf{Round (1--30)} \\
\quad Round (centered at 1) & $0.025^{***}\,(0.007)$ \\
\quad Medium $\times$ Round & $-0.020\,(0.010)$ \\
\quad High $\times$ Round   & $-0.040^{***}\,(0.010)$ \\
\addlinespace[4pt]
Task FE (domain) & Yes \\
Random intercept SD (user) & 1.563 \\
Observations & 4{,}500 \\
Users (groups) & 150 \\
\bottomrule
\end{tabular}

\vspace{0.5em}
\footnotesize
\emph{Notes.} Coefficients are log-odds; standard errors in parentheses. Model: GLMM (logit) with random intercept for subject ($1\,|\,\text{user\_id}$) and task fixed effects (domain). Baselines: density=\emph{Low}; task=\emph{cancer\_prediction}. Time is a continuous round index from 1--30, centered at 1. Significance:  $^{*}p{<}.05$, $^{**}p{<}.01$, $^{***}p{<}.001$.

\end{table}

\begin{table*}[h]
\centering
\caption{Delegation (logit, subject RE) with disclosure $\times$ performance: \textit{Cancer Prediction}. Baselines: No disclosure and Underperform.}
\label{tab:a1}
\begin{tabular}{lcccc}
\toprule
 & \textbf{Low density} & \textbf{Medium density} & \multicolumn{2}{c}{\textbf{High density}} \\
 & (No vs Partial) & (No vs Partial) & (No vs Partial) & (No vs Partial vs Full) \\
\midrule
\addlinespace[2pt]
\textbf{Disclosure (vs No)} \\
\quad Partial & $-0.114$\,(0.513) & $-0.317$\,(0.496) & $-0.466$\,(0.881) & $-0.470$\,(0.938) \\
\quad Full    & ---               & ---               & ---               & $-1.211$\,(0.893) \\
\addlinespace[4pt]
\textbf{Performance (vs Underperform)} \\
\quad Outperform & $0.853$\,(0.677) & $-0.551$\,(0.499) & $-1.852^{*}$\,(0.795) & $-1.898^{*}$\,(0.845) \\
\addlinespace[4pt]
\textbf{Interactions} \\
\quad Partial $\times$ Outperform & $-0.538$\,(2.034) & $0.634$\,(0.712) & $0.512$\,(0.934) & $0.522$\,(0.994) \\
\quad Full $\times$ Outperform    & ---               & ---               & ---               & $4.373^{***}$\,(1.263) \\
\addlinespace[6pt]
Time FE (round)  & Yes & Yes & Yes & Yes \\
Random intercept SD (user) & 1.759 & 1.307 & 1.126 & 1.221 \\
Observations & 720 & 830 & 960 & 1{,}150 \\
Users (groups) & 72 & 83 & 96 & 115 \\
AIC & 832.7 & 1{,}012.7 & 1{,}034.9 & 1{,}241.9 \\
\bottomrule
\end{tabular}

\vspace{0.5em}
\footnotesize
\emph{Notes.} Mixed-effects logit with random intercept for subject ($1|\texttt{user\_id}$) and time fixed effects (round). Coefficients are log-odds; standard errors in parentheses. Significance: $^{*}p{<}.05$, $^{**}p{<}.01$, $^{***}p{<}.001$.
\end{table*}

\begin{table*}[t]
\centering
\caption{Delegation (logit, subject RE) with disclosure $\times$ performance: \textit{Deceptive Hotel Reviews}. Baselines: No disclosure and Underperform.}
\label{tab:a2}
\begin{tabular}{lcccc}
\toprule
 & \textbf{Low density} & \textbf{Medium density} & \multicolumn{2}{c}{\textbf{High density}} \\
 & (No vs Partial) & (No vs Partial) & (No vs Partial) & (No vs Partial vs Full) \\
\midrule
\addlinespace[2pt]
\textbf{Disclosure (vs No)} \\
\quad Partial & $-0.630$\,(0.368) & $-0.257$\,(0.409) & $-1.334^{*}$\,(0.614) & $-1.374^{*}$\,(0.684) \\
\quad Full    & ---               & ---               & ---                   & $-2.114^{**}$\,(0.658) \\
\addlinespace[4pt]
\textbf{Performance (vs Underperform)} \\
\quad Outperform & $-0.996^{*}$\,(0.505) & $-0.226$\,(0.444) & $-2.471^{***}$\,(0.564) & $-2.560^{***}$\,(0.626) \\
\addlinespace[4pt]
\textbf{Interactions} \\
\quad Partial $\times$ Outperform & $1.161$\,(1.018) & $0.379$\,(0.713) & $1.115$\,(0.657) & $1.161$\,(0.733) \\
\quad Full $\times$ Outperform    & ---              & ---              & ---              & $4.802^{***}$\,(0.916) \\
\addlinespace[6pt]
Time FE (round)  & Yes & Yes & Yes & Yes \\
Random intercept SD (user) & 1.231 & 1.269 & 0.719 & 0.874 \\
Observations & 760 & 870 & 960 & 1{,}180 \\
Users (groups) & 76 & 87 & 96 & 118 \\
AIC & 952.4 & 1{,}084.2 & 1{,}113.8 & 1{,}363.6 \\
\bottomrule
\end{tabular}

\vspace{0.5em}
\footnotesize
\emph{Notes.} Mixed-effects logit with random intercept for subject ($1|\texttt{user\_id}$) and time fixed effects (round). Coefficients are log-odds; standard errors in parentheses. Significance: $^{*}p{<}.05$, $^{**}p{<}.01$, $^{***}p{<}.001$.
\end{table*}

\begin{table*}[t]
\centering
\caption{Delegation (logit, subject RE) with disclosure $\times$ performance: \textit{Loan Prediction}. Baselines: No disclosure and Underperform.}
\label{tab:a3}
\begin{tabular}{lcccc}
\toprule
 & \textbf{Low density} & \textbf{Medium density} & \multicolumn{2}{c}{\textbf{High density}} \\
 & (No vs Partial) & (No vs Partial) & (No vs Partial) & (No vs Partial vs Full) \\
\midrule
\addlinespace[2pt]
\textbf{Disclosure (vs No)} \\
\quad Partial & $-0.723$\,(0.430) & $-0.688$\,(0.484) & $-0.610$\,(0.470) & $-0.624$\,(0.522) \\
\quad Full    & ---               & ---               & ---               & $-1.225^{**}$\,(0.447) \\
\addlinespace[4pt]
\textbf{Performance (vs Underperform)} \\
\quad Outperform & $-0.251$\,(0.600) & $-1.012^{*}$\,(0.496) & $-1.339^{***}$\,(0.381) & $-1.372^{**}$\,(0.422) \\
\addlinespace[4pt]
\textbf{Interactions} \\
\quad Partial $\times$ Outperform & $1.306$\,(1.346) & $0.665$\,(0.710) & $0.325$\,(0.522) & $0.345$\,(0.579) \\
\quad Full $\times$ Outperform    & ---              & ---              & ---              & --- \\
\addlinespace[6pt]
Time FE (round)  & Yes & Yes & Yes & Yes \\
Random intercept SD (user) & 1.432 & 1.312 & 0.683 & 0.820 \\
Observations & 700 & 870 & 960 & 1{,}190 \\
Users (groups) & 70 & 87 & 96 & 119 \\
AIC & 854.3 & 1{,}064.8 & 1{,}164.0 & 1{,}440.5 \\
\bottomrule
\end{tabular}

\vspace{0.5em}
\footnotesize
\emph{Notes.} Mixed-effects logit with random intercept for subject ($1|\texttt{user\_id}$) and time fixed effects (round). Coefficients are log-odds; standard errors in parentheses. Significance: $^{*}p{<}.05$, $^{**}p{<}.01$, $^{***}p{<}.001$.
\end{table*}

\begin{table*}[t]
\centering
\caption{Always delegate (OLS with subject-clustered SEs). Dependent variable: indicator equals 1 if a participant always delegated within a task. Baseline disclosure is \emph{No}.}
\label{tab:a4}
\begin{tabular}{lcccc}
\toprule
 & \textbf{Low density} & \textbf{Medium density} & \textbf{High density} & \textbf{High density} \\
 & (No vs Partial) & (No vs Partial) & (No vs Partial) & (No vs Partial vs Full) \\
\midrule
\addlinespace[2pt]
Partial disclosure & $0.160^{*}$\,(0.075) & $-0.007$\,(0.057) & $-0.027$\,(0.029) & $-0.027$\,(0.029) \\
Full disclosure & --- & --- & --- & $0.236^{**}$\,(0.077) \\
\addlinespace[4pt]
\multicolumn{5}{l}{\textit{Task indicators (baseline: Cancer Prediction)}} \\
\quad Hotel Reviews & $-0.040$\,(0.043) & $-0.040$\,(0.040) & $0.000$\,(0.020) & $-0.023$\,(0.020) \\
\quad Loan Prediction & $0.020$\,(0.035) & $-0.040$\,(0.040) & $0.000$\,(0.025) & $-0.031$\,(0.027) \\
\addlinespace[6pt]
Task FE (domain) & Yes & Yes & Yes & Yes \\
Clustered SE (user) & Yes & Yes & Yes & Yes \\
Observations (N) & 300 & 300 & 300 & 390 \\
\bottomrule
\end{tabular}

\vspace{0.5em}
\footnotesize
\emph{Notes.} OLS with standard errors clustered at the subject level. Coefficients with standard errors in parentheses.
Significance: $^{*}p{<}.05$, $^{**}p{<}.01$, $^{***}p{<}.001$.
\end{table*}

\begin{table*}[t]
\centering
\caption{Mixed-effects logistic regression of delegation (log-odds). Random intercept by subject; task and time fixed effects included. Baseline treatment is \emph{No $\times$ Low}.}
\label{tab:a5}
\begin{tabular}{lcc}
\toprule
 & \textbf{No/Partial $\times$ Low/Med/High } & \textbf{No/Partial $\times$ Low/Med/High + Full $\times$ High} \\
\midrule
\addlinespace[2pt]
\textbf{Treatment (vs No$\times$Low)} & & \\
\quad No $\times$ Medium        & $-0.703^{*}$\,(0.329)  & $-0.712^{*}$\,(0.346) \\
\quad No $\times$ High          & $-1.371^{***}$\,(0.329) & $-1.395^{***}$\,(0.346) \\
\quad Partial $\times$ Low      & $0.438$\, (0.334)      & $0.454$\, (0.351) \\
\quad Partial $\times$ Medium   & $-0.784^{*}$\,(0.328)  & $-0.798^{*}$\,(0.345) \\
\quad Partial $\times$ High     & $-1.447^{***}$\,(0.326) & $-1.462^{***}$\,(0.343) \\
\quad Full $\times$ High        & ---                    & $0.083$\, (0.405) \\
\addlinespace[4pt]
\textbf{Controls (standardized)} & & \\
\quad Affinity for technology    & $-0.282^{**}$\,(0.103) & $-0.270^{**}$\,(0.104) \\
\quad AI literacy                & $-0.158$\, (0.111)     & $-0.174$\, (0.113) \\
\quad Risk                       & $-0.189$\, (0.113)     & $-0.216$\, (0.115) \\
\quad Task familiarity           & $-0.092^{*}$\,(0.046)  & $-0.113^{*}$\,(0.044) \\
\addlinespace[4pt]
Task fixed effects (domain) & Yes & Yes \\
Time fixed effects (round)  & Yes & Yes \\
Random intercept SD (user)  & 1.557 & 1.644 \\
Observations                & 9{,}000 & 9{,}900 \\
Users (groups)              & 300 & 330 \\
AIC                         & 9{,}763.9 & 10{,}597.4 \\
\bottomrule
\end{tabular}

\vspace{0.5em}
\footnotesize
\emph{Notes.} Coefficients are log-odds; robust standard errors in parentheses. Models: GLMM (logit) with random intercept for subject ($1|\text{user\_id}$), task fixed effects (domain), and time fixed effects (round). Controls are $z$-scored user means for affinity and AI literacy, $z$-scored risk, and a $z$-scored task-specific familiarity mapped to the row’s task. Significance: $^{*}p{<}.05$, $^{**}p{<}.01$, $^{***}p{<}.001$.
\end{table*}

\begin{table*}[t]
\centering
\caption{OLS regressions of total task coins. Cluster-robust SEs (subjects) in parentheses. Baseline treatment is \emph{No $\times$ Low}.}
\label{tab:a6}
\begin{tabular}{lcc}
\toprule
 & \textbf{(1) FE + controls} & \textbf{(2) + performance \& $AI$-share interactions} \\
\midrule
\addlinespace[2pt]
\textbf{Treatments (vs no\_low)} \\
\quad no\_medium        & $-39.129^{***}$\;(5.819) & $\phantom{-}4.034$\;(7.947) \\
\quad no\_high          & $-57.506^{***}$\;(6.985) & $12.860$\;(8.074) \\
\quad partial\_low      & $\phantom{-}35.792^{***}$\;(6.729) & $\phantom{-}9.033$\;(7.845) \\
\quad partial\_medium   & $-14.210^{**}$\;(5.424)  & $\phantom{-}2.620$\;(8.193) \\
\quad partial\_high     & $-50.719^{***}$\;(5.601) & $13.770$\;(7.578) \\
\quad full\_high        & $\phantom{-}35.014^{***}$\;(9.441) & --- \\
\addlinespace[4pt]
\textbf{Performance / Delegation} \\
\quad AI share                        & --- & $\phantom{-}72.308^{***}$\;(14.469) \\
\quad Own performance                 & --- & $123.545^{***}$\;(7.434) \\
\quad no\_medium $\times$ AI share    & --- & $-74.913^{***}$\;(20.191) \\
\quad no\_high $\times$ AI share      & --- & $-165.658^{***}$\;(20.865) \\
\quad partial\_low $\times$ AI share  & --- & $\phantom{-}43.889^{*}$\;(17.896) \\
\quad partial\_medium $\times$ AI share & --- & $-21.433$\;(20.685) \\
\quad partial\_high $\times$ AI share & --- & $-150.888^{***}$\;(19.225) \\
\addlinespace[4pt]
\textbf{Controls (z-scored)} \\
\quad Affinity for technology & $-4.482^{*}$\;(2.130) & $-2.424$\;(1.332) \\
\quad AI literacy             & $\phantom{-}4.441^{*}$\;(2.184) & $\phantom{-}1.761$\;(1.572) \\
\quad Risk                    & $-6.738^{**}$\;(2.249) & $-0.385$\;(1.400) \\
\quad Task familiarity        & $-3.975$\;(2.237) & $\phantom{-}0.351$\;(1.412) \\
\addlinespace[4pt]
Task FE (domain) & Yes & Yes \\
SEs & Clustered by user & Clustered by user \\
\bottomrule
\end{tabular}
\vspace{0.5em}

\footnotesize
\emph{Notes.} Column (1): OLS with treatment fixed effects and controls. Column (2): adds own performance and $AI$-delegation share with treatment\,$\times$\,AI-share interactions; the full\_high cell is excluded in this six-cell specification. Significance: $^{*}p{<}.05$, $^{**}p{<}.01$, $^{***}p{<}.001$.
\end{table*}

\clearpage

\begin{table}[t]
\centering
\renewcommand{\arraystretch}{1.3} 
\caption{Post-hoc pairwise comparisons using Dunn test with Bonferroni correction for the number of AI hovered per condition ($^{*}p{<}.05$, $^{**}p{<}.01$, $^{***}p{<}.001$).}
\begin{tabular}{lcccc}
\toprule
\textbf{Comparison} & \textbf{Z} & \textbf{p\_raw} & \textbf{p\_adj}\\
\midrule
full\_high - partial\_high & 3.0829 & .0020 & .0123*  \\
full\_high - partial\_low & -0.7411 & .4586 & 1.0000  \\
partial\_high - partial\_low & -4.3568 & .0000 & .0001*** \\
full\_high - partial\_medium & 0.3352 & .7375 & 1.0000 \\
partial\_high - partial\_medium & -3.1553 & .0016 & .0096**  \\
partial\_low - partial\_medium & 1.2288 & .2192 & 1.0000 \\
\bottomrule
\end{tabular}
\label{tab:hovered_ais}
\end{table}

To obtain further insights about participants' behaviors, we explored the average numbers of AIs they hovered during the study, considering all the \partd{} conditions and the \fulld{} benchmark (see Table \ref{tab:hovered_ais} in the Appendix). Preliminary analysis indicated that the data were not normally distributed, as demonstrated by the Shapiro-Wilk normality test  ($W = 0.916$, $p < .001$). Therefore, we employed the non-parametric Kruskal-Wallis test, which revealed significant differences between conditions ($\chi^2 = 21.5$, $df = 3$, $p < .001$). 
Based on this result, post-hoc pairwise comparisons were conducted using the Dunn test with Bonferroni correction for multiple comparisons. 
The comparisons showed that \textit{partial\_high} had significantly higher AIs hovering from \textit{partial\_low} 
($Z = 4.36$, $p_{adj} < .001$) and from \textit{partial\_medium} 
($Z = 3.16$, $p_{adj} = .0096$). 
Additionally, \textit{full\_high} conditions had significantly less AIs hovered from \textit{partial\_high} 
($Z = -3.08$, $p_{adj} = .0123$).

\end{document}

%% file: 1_New_Introduction.tex
\section{Introduction}
\label{sec:introduction}

\revision{
AI systems are increasingly integrated into consumer-facing applications, yet their complexity often obscures critical quality dimensions such as reliability, safety, and fairness from end-users \citep{obermeyer2019dissecting,nwafor2024enhancing,jin2018artificial}. While these systems can appear highly accurate, they may still make costly mistakes or exhibit hidden defects \citep{cash2025quantifying,vasconcelos2023explanations,rathi2025humans,choudhury2024large,steyvers2025large,simhi2025trust,li2025confidence}. From an HCI perspective, this opacity creates significant challenges for trust calibration, appropriate reliance, and user agency. Users often lack the ability to verify system quality either before or after use \citep{biermann2022algorithmic,fok2024search}, which can lead to misaligned mental models and poor decision-making.
}

\revision{In the field of economics, this is traditionally framed as a \enquote{market for lemons} problem, where uncertainty about product quality drives inefficient outcomes \citep{akerlof1978market}. In the context of human-AI interaction, this translates into information asymmetry between designers or deployers of AI systems and their users, where users cannot easily assess whether an AI system is fit for their task. This can lead to sub-optimal user decisions not only on whether to adopt an AI system, but also on which AI system to adopt, given the presence of alternatives. While mechanisms such as reputation systems, third-party certifications, and disclosure labels (e.g., Apple’s privacy “nutrition” labels) aim to bridge this gap \citep{grossman1981informational,scoccia2022empirical,shapiro1983premiums,spence1977consumer}, they often fail in practice due to selective reporting, high cognitive load, or strategic complexity that undermines transparency \citep{liang2024systematic,ali2023honesty,jin2022complex}.} 
\revision{Recent regulatory efforts, such as the EU AI Act, mandate transparency, but standards remain fragmented and lack empirical grounding in user behavior \citep{kaminski2021algorithmic,van2022artificial}. Evidence suggests that current disclosure practices (e.g., model cards) vary widely in informativeness and usability \citep{liang2024systematic,ribeiro2020beyond}. Importantly, providers of these systems are generally incentivized to reveal favorable but hide unfavorable information \citep{rayo2010optimal,jin2021no}, or use unnecessary complexity to shroud information \citep{jin2022complex}, which constraints the usefulness of voluntary transparency efforts. Addressing this requires a regulatory framework that is strategy-proof, such that suppliers cannot ``game'' the system through partial information revelation \citep{raji2020closing}.}

\revision{In this paper, we provide the first experimental evidence on how information asymmetries (modeled through disclosure design) and the relative density of low-quality AI systems (also referred to as \textit{lemons}) in the available pool for users, influence their adoption and reliance on AI systems. To this end, we simulate an interactive decision-making environment in which 
participants complete three different tasks across 10 rounds, in each of which they can decide to use assistance from an AI system of their choice from a pool of alternatives. Our controlled study follows a 3 (lemon density: \lowden{} vs \medden{} vs \highden{}) $\times$ 2 (information disclosure: \nod{} vs \partd{}) between-subjects design, with an additional benchmark condition of \highden{} with \fulld{}.} The overall quality of an AI system in our study is represented by an \textit{accuracy score} and a \textit{data quality} score, where the latter represents the AI system's generalizability.\footnote{Note that our technical operationalization of AI system quality does not consider other relevant dimensions of system quality such as robustness, fairness, uncertainty, or alignment with human values---where data quality remains a fundamental enabler. 
This operationalization is a principled simplification that necessarily excludes other important dimensions of AI system quality that fall outside the scope of our study, while remaining externally valid. Several dimensions of AI system quality in the real world are often encapsulated in composite metrics that are conveyed to consumers to help shape their mental models of the systems.}
In the \partd{} conditions, participants only observe the accuracy score, whereas the \fulld{} conditions reveal both accuracy and data quality. \revision{This design allows us to examine how users interpret and act on these cues under varying conditions of low-quality AI system density, as illustrated in Fig. \ref{fig:teaser}. We address the following research questions:}

\begin{framed}
\begin{itemize}[leftmargin=*]
    \item \textbf{RQ1}: How do information asymmetries about AI system capabilities affect \revision{users' adoption of AI} and market outcomes in the presence of low-quality AI systems?

    \item \textbf{RQ2}: How do different information disclosure requirements about AI system capabilities impact \revision{user behavior}, market outcomes, and reliance on \revision{low-quality AI systems}?
\end{itemize}    
\end{framed}


\revision{We contribute a novel experimental framework that adapts the classic economics theory of “market for lemons” \cite{akerlof1978market}  to human-AI interaction, operationalizing lemon density and disclosure strategies to systematically study how information asymmetry and uncertainty shape user reliance and decision-making.}
\revision{Our findings show that users are sensitive to the presence of low-quality AI systems when no disclosure is provided (\nod{}), but they struggle to calibrate their reliance effectively over time. When the proportion of low-quality systems is low, participants tend to underutilize AI assistance, missing opportunities to improve performance. Conversely, when low-quality systems are prevalent, they over-rely on AI, delegating tasks even when it is detrimental. While some learning occurs across rounds, its magnitude is small, suggesting persistent challenges in forming accurate mental models of system quality.}

\revision{We found that introducing partial disclosure (i.e., revealing only AI system accuracy) significantly improves decision-making efficiency. Participants leverage these incomplete signals to avoid low-quality systems at high rates, consistent with adaptive trust calibration. This leads to substantial performance gains in \lowden{} and \medden{} conditions, though benefits diminish when low-quality systems dominate (i.e., in the \highden{} conditions). Interestingly, the participants' ability to interpret and act on partial disclosure cues declines as the density of poor-quality systems increases, highlighting the limits of simple transparency mechanisms. 
Importantly, partial disclosure does not change overall delegation rates or the proportion of participants who fail to optimize their choices, indicating that information disclosure primarily improves the quality of delegated decisions rather than the quantity. In fact, the effect is strong enough to offset a doubling of low-quality systems in the market.
In contrast, our full disclosure benchmark (\fulld{} with \highden{})--where users see both accuracy and data quality--reveals a different challenge: even when fully informed, participants exhibit inefficiently low AI adoption. While they successfully avoid low-quality AI systems, only about 58\% of predictions are delegated to systems that are correct with a probability of 90\%, resulting in average losses of around 20\% compared to full delegation. This suggests that a greater degree of information disclosure does not automatically translate into better outcomes, and that other factors (e.g.,  risk aversion) may lead users to under-utilize AI systems even when it is objectively beneficial. We also found that the average performance in \partd{} with \lowden{} matches that of \fulld{} with \highden{}, underscoring the importance of designing disclosure for actionability rather than completeness.}

\revision{While disclosure mitigates information asymmetry, it does not eliminate other cognitive and behavioral barriers such as bounded rationality~\cite{kaur2024interpretability}, risk aversion, and trust calibration~\cite{mehrotra2024systematic} challenges. This nuance shows that system transparency is necessary but not sufficient to foster appropriate adoption and reliance on AI systems. 
Apart from important design implications for human-AI interaction, our work has direct implications for emerging standards such as the EU AI Act, where enforceable yet lightweight disclosure rules may be more effective than complex, fully detailed reporting in promoting appropriate reliance on AI systems.}

%% file: 2_RelatedWork.tex
\section{Background and Related Work}
\label{sec:related_work}

\subsection{Human Beliefs and Reliance on AI Systems}


Behaviorally, our paper relates to the literature at the intersection of user information, user beliefs, and AI utilization. \citet{biermann2022algorithmic} provides experimental evidence that people struggle ex ante and ex post to verify the prediction quality of algorithms, which can even be exacerbated by explanations. Users learn over time, which has implications for small but not for large markets. In general, humans use accuracy signals to update their beliefs and adjust reliance behavior, but tend to calibrate imperfectly \citep{he2023stated,erlei2024understanding,biswas2026belief}. Furthermore, additional information may also fail to improve human reliance if they cannot interpret the signals correctly or experience information overload \citep{poursabzi2021manipulating}. 
Research on over-reliance has argued that confirmation bias, anchoring bias and base neglect may explain why people sometimes apparently under-weigh observable shortcomings of AI decision making systems \citep{mosier2017automation,nourani2021anchoring,rastogi2022deciding,mikhaylova2025bayesian,gadiraju2025enterprising}. In the context of explainable AI (XAI), users have exhibited a variety of cognitive biases related to the processing of information signals, such as representative and availability biases, choice overload, or relative inattention to false positives \citep{bertrand2022cognitive}. In our experiment, we rely on information labels as partial signals that allow rational users to improve their decision-making but cannot rule out false negatives (i.e. accidentally approaching a lemon). False positives (wrongly identifying a high-quality AI system as a lemon) do not play a role \citep[see also][]{goodyear2017fmri,kocielnik2019will}. The main reason is that belief updating under information asymmetries across different tasks is already cognitively demanding. Here, focusing on false negatives is not only externally more valid for regulatory purposes, but also streamlines the cognitive effort of participants to the processes relevant to our main research questions. \citet{agarwal2023combining} analyze how radiologists with access to AI assistants update their beliefs and concurrent diagnostic choices when the AI signal is (i) certain or (ii) uncertain. They find improvements in decision quality under certainty, while radiologists fail to update under uncertainty. In \citet{reverberi2022experimental}, endoscopists were able to rationally integrate AI advice into their diagnostic process. Note that in these cases, experts interact with expert systems with verified, although on a case-by-case basis uncertain, prediction quality. Thus, there is individual-level evidence that (expert) users can rationally integrate AI-generated advice following Bayesian principles, which may be inhibited by uncertainty in the choice environment. 

\revision{Researchers have 
studied how user mental models of AI systems, their experiences, expectations, and various AI metaphors can influence and shape experiential and performance outcomes in human-AI collaboration and decision-making contexts~\cite{bansal2019beyond,khadpe2020conceptual,jacovi2021formalizing}. Recently, \citet{gadiraju2025enterprising} synthesized various human, task, and AI system factors empirically shown to shape outcomes in human-AI collaboration.}
\revision{Prior HCI work on trust calibration and transparency has largely focused on static disclosure artifacts (e.g., model cards~\cite{mitchell2019model}, impact assessment reports~\cite{bogucka2024ai}) or interpretability cues in isolation. Our work introduces a behavioral economics lens to HCI, operationalizing the \enquote{market for lemons} theory as an experimental framework to study AI adoption under varying information asymmetries and varying densities of low-quality AI systems. This is a novel theoretical integration that connects economic models of uncertainty with human-centered design challenges.}

\subsection{Information Asymmetry in Human-AI Collaboration: Technical Solutions and Regulatory Frameworks}


The asymmetry of information between AI systems and end-users is particularly evident in black-box models, whose decision-making processes are often opaque and difficult to interpret \cite{longoXAIManifesto24}. This lack of transparency can undermine trust, reduce accountability, and hinder users' ability to evaluate the reliability of AI outputs \cite{burrell2016how, yang2023survey}. Explainable AI (XAI) approaches aim to address this gap by providing intelligible explanations that help users understand the factors driving model predictions \cite{doshi2017towards, miller2019explanation,AliXAIReview24,He2025ConversationalXAIOnDemand}. XAI literature spans different approaches that have been proposed, ranging from local \cite{Lai2019ExampleBasedFeatureBasedAIPredictions,Chen2023RelianceExampleBasedFeatureBased,Bove2022ContextualizationExplortationExampleBasedLocalFeatureImportance,Lai2023SelectiveExplanations} to hybrid \cite{Anik2021DataCentric,Bhattacharya2023DirectiveExplanations,Bhattacharya2024EXMOS,Bhattacharya2024ModelSteeringSystem,Szymanski2024HealthDashboardDataCentric,Cau2025CuriosityTraits} explanations, which involve the integration of data-centric explanations with model-centric ones to promote improved human-AI decision-making and appropriate reliance. 

An alternative approach is the use of inherently interpretable, or \enquote{glass-box} models \cite{rudin2019stop}. While black-box models often achieve higher predictive accuracy, glass-box models enhance transparency, enabling users to detect and correct data quality issues such as biases, errors, or mislabeling \cite{rudin2019stop, cheng2025comprehensive}. High-quality, well-labeled, and representative datasets improve the performance of both model types, but the interpretability of glass-box models makes it easier to identify and address problems in the data \cite{cheng2025comprehensive, yang2023survey}. Hybrid approaches that combine accurate black-box models with XAI techniques can offer a balanced solution, maintaining predictive performance while providing users with interpretable insights, which could reduce information asymmetry. In this paper, we represent this real-world context where users can interact with AI systems with varying access to information about indicators of their quality.

Addressing the growing concerns and downstream consequences of information asymmetry in human-AI collaboration, regulatory frameworks in the European Union have been designed to enforce transparency and accountability. The EU AI Act introduces risk-based obligations, requiring providers to disclose when exactly users are interacting with AI systems (EU AI Act, Article 50).\footnote{\url{https://artificialintelligenceact.eu/article/50/} [last accessed on: 11 September 2025]} It mandates the explicit labeling of AI-generated content and encourages explainability in high-risk systems. As a complement to this, the General Data Protection Regulation (GDPR) ensures that individuals are informed about how their personal data is processed by AI, granting rights to explanation and human intervention in automated decision-making (GDPR, Articles 13–22).\footnote{\url{https://gdpr-info.eu/art-13-gdpr/} [last accessed on: 11 September 2025]} In contrast, the Digital Services Act (DSA)\footnote{\url{https://digital-strategy.ec.europa.eu/en/policies/digital-services-act-package} [last accessed on: 11 September 2025]} targets online platforms, requiring transparency in algorithmic content moderation and recommendation systems, and obligating platforms to disclose AI involvement in personalized content and ads. These regulations collectively contribute to the attempts at reducing information asymmetry in human-AI collaboration and aim to promote responsible AI deployment across different domains \cite{balasubramaniam2023transparency,cheong2024transparency}. Despite these efforts, complexity of the global AI market and benchmarking practices have meant that users continue to interact with and rely on AI systems with limited access to information about their quality. Our work contributes to understanding the impact of regulatory institutions on AI adoption, user behavior, and market outcomes---particularly when low-quality AI systems (\textit{lemons}) are prevalent in the market. \revision{Our experimental framework is both novel and extensible. Future studies can vary other factors such as task complexity, stakes, fairness signals, or robustness indicators using the same market-based paradigm.}  


\subsection{Disclosure}
Most research surrounding the desirableness and efficacy of disclosure institutions comes from economics, following the classic ``unraveling'' logic. When quality is verifiable and disclosure is costless, suppliers generally voluntarily reveal quality (except the worst anti-social types) \citep{grossman1981informational,milgrom1981goodnews}. In practice, unraveling is frequently incomplete. Sellers face strategic incentives to disclose selectively, delay or bury unfavorable information, or increase complexity to shroud key attributes \citep{rayo2010optimal,jin2021no,gabaix2006shrouded,ellison2009search,jin2022complex}. These frictions are central to AI markets where suppliers can highlight favorable benchmarks while under-reporting limitations, safety risks, or domain shift -- precisely the conditions our design emulates with partially informative labels. 

A substantial body of empirical literature shows that third-party verification and mandatory disclosure can (partially) discipline markets with hidden qualities. In restaurants, making hygiene report cards salient improves practices and shifts demand toward higher-grade establishments \citep{jin2003restaurant,jin2009reputation}. In healthcare, public “report cards” affect provider behavior and consumer choice, but also induced risk selection, highlighting design trade-offs, such as the scope of suppliers to ``game'' the system \citep{dranove2003reportcards}. Publicized plan ratings change enrollment in health insurance markets \citep{jin2006healthplans}. In consumer goods, voluntary disclosure regarding nutrition labeling is asymmetric (good types disclose, bad types often do not), whereas mandatory labeling shifts demand and improves welfare by making negative attributes visible \citep{ippolito1990cereal,mathios2000salad,bollinger2011calorie}. Broadly, the evidence points towards a positive effect of well-designed disclosure and certification institutions on selection and market-wide product quality. However, there is a lot of heterogeneity, specifically with regards to signal type, credibility, domain context and scope. Most evidence comes from experience goods that are comparatively simple to regulate.

Disclosure’s efficacy also depends on how information is presented and processed. For example, salience and simplicity can determine whether consumers are able to exploit signals \citep{chetty2009salience,bollinger2011calorie}. Particularly in the AI context under market information asymmetries, simplicity is a complex problem, given the Bayesian nature of the involved information and learning process. Certification experiments show sizable premia for verified quality and reductions in “lemons” risk when credible third parties attest to product characteristics \revision{\citep{jin2006certification}}. When markets rely solely on reputational mechanisms, disclosure may be too noisy or manipulable to separate types in thick, anonymous environments \citep{nosko2015limits}. This directly maps to AI markets, where voluntary “model cards” or benchmark disclosures can be selectively curated, and where strategic non-disclosure or obfuscation is profitable in the absence of enforceable, strategy-proof \textit{and} behaviorally valid rules \citep{rayo2010optimal,jin2021no,jin2022complex}.

This study leverages these insights in three ways. First, we test a \partd{} regime that is realistically noisy. It represents the empirically observed gap between theoretical unraveling and real-world selective disclosure. Second, we benchmark against \fulld{} to isolate the upper bound on efficiency when information asymmetries have been solved. Third, we vary the complexity of the market by manipulating the density of lemons, which also allows us to make sharp behavioral predictions.

%% file: 3_Method.tex
\section{Method}
\label{sec:methodology}
This section outlines how we developed the user study to assess the ``market for lemons'' problem for AI adoption, starting with an overview of the lemon market we simulate in our study, the selection of different types of tasks, the AI system pool, instances, and the design of the AI-assisted interface. 

\subsection{\revision{\textbf{Simulation of The Lemon Market}}}

\revision{Our study aims to} 
simulate a market environment in which consumers demand some AI good (e.g., an AI system that can aid a decision-making task) but are ex ante uncertain about the system's quality. Consumers observe the quality after usage, but operate in a large market, such that their experiences do not meaningfully affect their search set, and hence do not reduce uncertainty for future purchases. 
There is a single posted price 
because low‑quality sellers mimic high‑quality ones. 
\revision{Our simulation of the lemon market is motivated by the increasing reliance of users on AI systems to support decisions today (e.g., choosing a resume screener, a grammar checking tool, or a writing aid). Yet users often cannot tell whether an AI system is broadly reliable before they commit to it. Benchmark accuracy labels can look impressive, but may not translate to real‑world generalizability and mislead users (e.g., due to biased data or domain shifts). Thus, simulating this market can allow us to systematically study how different levels of disclosure shape user adoption of AI systems in a market where some AI systems are truly high quality (i.e., \textit{peaches}) and others are low quality (i.e., \textit{lemons}).}

Formally, there are two types of AI systems $\theta \in \{H,L\}$ characterized by quality $Q = (\alpha, g)$ that differ in terms of measurable accuracy $\alpha \in [0,1]$ and generalizability due to data quality $g \in [0,1]$. High-quality AI systems (i.e., \textit{peaches}) $H$ always exhibit $Q_H = (\alpha_H, g_H)$, and low-quality AI systems (i.e., \textit{lemons}) exhibit either $Q_L = (\alpha_H, g_L)$ or $Q_L = (\alpha_L, g_L)$ with $g_H > g_L$, and $\alpha_H > \alpha_L$. Here, we assume that lemons always exhibit poor generalizability ($g_L$) due to fundamental data issues such as dataset biases or sampling issues, overfitting to training distributions, or domain shifts grounded in real-world considerations. However, some lemons can exhibit high accuracy ($\alpha_H$) in external tests. For lemons, there is heterogeneity in measurable AI accuracy from external testing, e.g., through benchmark platforms. Thus, lemons exhibiting high accuracy scores have incentives to disclose these scores voluntarily. Peaches, on the other hand, always exhibit high accuracy scores and are not plagued by data-related issues. Therefore, accuracy is only partially informative. Consumers earn some gross surplus $u_H > 0$ from using a high-quality AI system, but experience a cost $u_L < 0$ when purchasing a low-quality AI system.\footnote{Consumer surplus in economics is the monetary gain that consumers make when they purchase a product or service for a price that is less than the highest price they are willing to pay \cite{marshall2013principles}.} The uniform market price is $p$. Depending on the experimental conditions, consumers can (not) observe the share of lemons in the market $\lambda \in [0,1]$ and receive a signal $s$ about the quality of each AI system based on three information regimes. In the \nod{} conditions, consumers condition their choices based on what they learn about AI system quality in the market across multiple trials. In the \partd{} conditions, consumers observe a signal $s \in {0,1}$ in the form of an accuracy badge label that represents the measurable accuracy (either $\alpha_H$ or $\alpha_L$). If $s = 1$, the product has high accuracy. Consumers always know that there is a probability $\gamma = 1$\footnote{Results are unchanged for $\gamma < 1 \, \text{as long as} \, \gamma > \beta$.} for $s = 1$ if $\theta = \textbf{H}$, and a probability $0 < \beta < \gamma$ for $s = 1$ if $\theta = \textbf{L}$. Formally, \textbf{Pr}$[s = 1 | H] = \gamma$, \textbf{Pr}$[s = 1 | L] = \beta$. Then, consumers who observe the signal $s$ update their beliefs that any AI is of high quality using Bayes' rule after seeing the accuracy label:
\begin{equation*}
    \begin{aligned}
    \textbf{Pr}[\theta = H | s = 1] &= \frac{(1-\lambda)\gamma}{(1-\lambda)\gamma + \lambda \beta}, \\[6pt]
    \vee \quad \textbf{Pr}[\theta = H | s = 0] &= \frac{(1-\lambda)(1-\gamma)}{(1-\lambda)(1-\gamma) + \lambda(1-\beta)}.
    \end{aligned}
\end{equation*}

\noindent
A key feature of our signaling structure is that while the $s = 1$ signal is noisy, the $s = 0$ signal is perfectly informative and always reveals a lemon. This simplifies exposure for consumers and gives a sharp behavioral prediction whereby rational consumers never purchase a low-accuracy AI product. It is the minimal requirement for partial disclosure regulations. Finally, the \textbf{full} disclosure reveals $\theta$ perfectly, consumers are fully informed. 
\revision{To elucidate, even a simple badge (that supports partial disclosure) can reduce harmful AI adoption if low‑accuracy labels reliably flag lemons. But unless generalizability is disclosed, some lemons will still pass as high accuracy. Full disclosure solves adverse selection by aligning user decisions with true quality.}

\noindent
We make five \revision{important} assumptions. One, consumers are risk-neutral and maximize expected monetary surplus $E[u_{\theta}] - p$. Two, all agents know $\lambda$, $u_H, u_L, p, \gamma, \beta$.\footnote{In \partd{} and \fulld{}, $\lambda$ is observable. In \nod{}, subjects cannot observe $\lambda$, but learn it over the course of 30 rounds. Here, the prediction hinges on consumer (i) exploration of the AI market and (b) learning.} Three, the search space is constant across rounds, each purchase is independent (large market). Fourth, lemon sellers post the same price as high-quality sellers. This rules out, for example, side-payments, or incentives for endogenous segmentation. Fifth, disclosure changes neither production cost, nor prices in the short run, $p$ is fixed throughout the experiment.

\textbf{Prediction.} The risk-neutral consumer buys if $EU(x) = Pr[\theta = H | x]u_H + Pr[\theta = L | x]u_L - p \geq 0$ where $x \in \{\emptyset, s = 1, s = 0, \theta = H, \theta = L\}$. In \nod{}, they can only condition their decision on the (learned) prior $\lambda$: $EU(\emptyset) = (1-\lambda)u_H + \lambda u_L - p$. In \partd{}, $EU(s = 1) = [\frac{(1-\lambda)\gamma}{(1-\lambda)\gamma + \lambda \beta}]u_H + [1- \frac{(1-\lambda)\gamma}{(1-\lambda)\gamma + \lambda \beta}]u_L - p$ or $EU(s = 0) = [\frac{(1-\lambda)(1-\gamma)}{(1-\lambda)(1-\gamma) + \lambda(1-\beta)}]u_H + [1 - \frac{(1-\lambda)(1-\gamma)}{(1-\lambda)(1-\gamma) + \lambda(1-\beta)}]u_L - p$. In \fulld{}, we get $EU(H) = u_H - p > EU(L) = u_L - p$. Finally, we assume $p < u_H$ such that it makes sense for consumers to enter the market for AI products. Using this simple decision model, we set our experimental parameters such that, on average given priorly published behavioral data from user studies (see section Experimental Parameters below), $(1-\lambda)u_H + \lambda u_L > p$ in \lowden{}, $(1-\lambda)u_H + \lambda u_L < p < [\frac{(1-\lambda)\gamma}{(1-\lambda)\gamma + \lambda \beta}]u_H + [1- \frac{(1-\lambda)\gamma}{(1-\lambda)\gamma + \lambda \beta}]u_L$ in \medden{}, and $[\frac{(1-\lambda)\gamma}{(1-\lambda)\gamma + \lambda \beta}]u_H + [1- \frac{(1-\lambda)\gamma}{(1-\lambda)\gamma + \lambda \beta}]u_L < p < u_H$ in \highden{}.

\noindent
Note that we focus on a short-run analysis with exogenous prices.  Our main focus is on consumer behavior, and the effectiveness of different disclosure regimes, rather than seller behavior. If price were endogenous in a competitive market with free entry, high quality sellers need $p \geq c_H$, and lemons need $p \geq c_L$ where $c_H > c_L$. Consumers cannot separate types, and are therefore willing to pay up to $p = E(x)$. For instance, in \textbf{no} disclosure, their willingness to pay (WTP)\footnote{WTP is a core concept in economics that represents the maximum amount of money an individual is prepared to spend to acquire a specific good or service.} is $E(\emptyset) = (1-\lambda)u_H + \lambda u_L$. As $\lambda$ rises, $E(x)$ and hence the market price $p$ falls, where eventually $p < c_H$ and high-quality sellers exit the market, once again increasing $\lambda$, leading to the well-known ``death spiral'' \citep{akerlof1978market}. Full disclosure solves that problem by driving lemons prices downwards to $p = c_L$.

\subsection{Experimental Parameters}
In our experiment, participants choose between themselves and an AI system. Instead of setting a fixed exogenous price, we model the price as the average opportunity costs of relying on an AI system, i.e., average subject performance. Prior research suggests user accuracy of 50\%--60\% for all three chosen tasks \cite{he2023stated,Ott2011DeceptiveHotelReviews,Cabrera2023DeceptiveHotelReviews,Qu2025DeceptiveHotelReviews,Cao2024UncertaintyPresentation}. Following that, we assume an average prediction accuracy of 55\%. We set average lemon accuracy to 15\% and average accuracy of a high-quality AI system to 90\%. Participants earn 30 Coins (\$0.1) per correct prediction, and 0 otherwise. This gives $p = 16.5$, $u_H = 27$ and $u_L = 4.5$. The publicly known probability of a lemon to exhibit high accuracy is $\beta = \frac{1}{3}$. The risk-neutral expected payoffs (assuming that participants correctly use signal $s$ and, in \nod{}, learn the prior $\lambda$) are shown in the Table \ref{tab:pay} below.

\begin{table}[h]
\caption{Expected Average AI-Market Payoffs Between Conditions}
\label{tab:pay}
\footnotesize
\begin{tabular}{lccc}
\toprule
& {\lowden} & {\medden} & {\highden} \\
\midrule
\nod{}      & 4.5       & -3        & -8.25      \\
\partd{}    & 8.79      & 3         & -4.875     \\
\fulld{}    & {---}     & {---}     & 12         \\
\bottomrule
\end{tabular}
\end{table}

The expected payoff depends on both density and disclosure institution, where under \medden{}, partial disclosure flips the expected payoff from negative to positive. In \highden{}, relying on the AI market is economically harmful in expectation, except if information asymmetries are fully resolved. Our predictions follow these payoffs. Higher expected returns increase the share of subjects who -- if they efficiently utilize the information signals -- rely on the AI market. Hence, delegation increases along the disclosure institutions and decreases along the density condition. The majority of subjects consistently use the AI market in \lowden{} irrespective of the disclosure condition, whereas in \medden{}, this pattern only emerges in \partd{}. For \highden{} we predict very low rates of AI adoption, with higher delegation shares in \partd{} than \nod{}.


 

\revision{\subsection{Hypotheses}}
\revision{Following the conditions described above, we derive four hypotheses that we aim to test in our study:
\begin{itemize}
    \item \textbf{H1} (\textit{effect of disclosure}): AI adoption will increase with the level of disclosure (i.e., AI adoption in the experimental conditions will follow: \fulld{} > \partd{} > \nod{}.)
    \item \textbf{H2} (\textit{effect of lemon density}): AI adoption will decrease as the share of lemons in the market increases (i.e., AI adoption in the experimental conditions will follow: \lowden{} > \medden{} > \highden{}.)
    \item \textbf{H3} (\textit{interaction}): AI adoption will vary across the different lemon density and disclosure conditions as follows:
    \begin{itemize}
        \item [\textbf{H3a}:] In \lowden{} conditions, participants will consistently delegate to AI across all disclosure conditions.
        \item [\textbf{H3b}:] In \medden{} conditions, participants will delegate only under \partd{} or \fulld{}.
        \item [\textbf{H3c}:] In \highden{} conditions, participants will delegate only under \fulld{}.
    \end{itemize}
    \item \textbf{H4} (\textit{decision-making efficiency}): When disclosure is available, participants will use the information efficiently (e.g., avoid low-accuracy badges), leading to higher decision-efficiency compared to \nod{}.
\end{itemize}}

\subsection{Tasks Selection}
\label{sec:task_selection}
To gain a broader understanding of how information asymmetry and disclosure impact AI adoption in a lemon market, we selected three different tasks with varying data modality 
that are widely used in the Human-Centered AI (HCAI) literature: skin cancer prediction, loan approval, and deceptive review detection. \revision{Note that all task data are appropriately anonymized and available publicly.}

\subsubsection{Skin cancer prediction}
For image data, we opted for the \textit{ISIC
2018 challenge dataset}\footnote{\url{https://datasetninja.com/isic-challenge-2018}} given its suitability for AI-assisted decisions as in previous work \cite{Tschandl2019SkinCancerDetectionISIC,Tschandl2020SkinCancerDetectionISIC,Barata2023SkinCancerDetectionISIC,Chanda2024SkinCancerDetectionISIC,Cao2024UncertaintyPresentation,Yan2025SkinCancerDetectionISIC}. 
Specifically, we set up a binary skin cancer prediction task 
where participants need to decide, given a skin image, whether it appears as benign or shows signs of malignant skin cancer (e.g., a general skin cancer).

\subsubsection{Loan approval} 
For tabular data, we selected the \textit{loan prediction problem dataset}\footnote{\url{https://www.kaggle.com/datasets/altruistdelhite04/loan-prediction-problem-dataset}} as a testbed, given its utility in several previous works in AI-assisted decisions
\cite{Binns2018Loan,Green2019Loan,Gomez2020ViCELoan,Chromik2021Loan,VanBerkel2021Loan,he2023stated,Esfahan2024LoanPrediction,He2025ConversationalXAIOnDemand}. 
In this task, participants are asked to decide whether to accept or reject a loan application based on twelve attributes of an applicant (e.g., employment status, education level, credit history, etc.). To avoid creating an ambiguity effect as shown in prior literature~\cite{eickhoff2018cognitive}, we removed the Loan-ID attribute, as it provides little information for decision-making---resulting in eleven attributes. 

\subsubsection{Deceptive review detection}
For text data, we chose \textit{deceptive hotel reviews dataset}\footnote{\url{https://myleott.com/op-spam.html}} since it has been used in previous studies of human-AI collaboration
\cite{Ott2011DeceptiveHotelReviews,Lai2020TextDeceptiveReview,Arora2022DeceptiveHotelReviews,Cabrera2023DeceptiveHotelReviews,He2024DeceptiveHotelReviews,Qu2025DeceptiveHotelReviews}. 
The goal of the task is to determine whether a review is \enquote{genuine}, hence written by someone who stayed at the hotel, or \enquote{deceptive}, so written by someone who has not. As done in previous work \cite{Cabrera2023DeceptiveHotelReviews}, we selected genuine reviews from online sites such as TripAdvisor that involve positive polarity. Instead, deceptive reviews were collected from Amazon Mechanical Turk workers. 

\subsection{Instances and AI System Selection}
\label{sec:instance_selection}

For each task, we selected ten instances, ensuring a balance between the binary true classes, with five for the positive class (i.e., benign, accept, or genuine) and five for the negative class (i.e., malignant, reject, or deceptive). 
For each instance, we instantiated ten simulated AI systems with different fixed combinations of accuracy (low = 65\%, or high = 90\%) and data quality (low or high). 
We adjusted the accuracy values by adding a random noise with a range of 0-3\% to the baseline values to simulate a refresh effect on the AI pool after each trial completion, further randomizing the order of the AI systems. Thus, we clearly indicate to the participants that the pool of AI systems is renewed after each trial. 
This design has several advantages. One, we ensure that subjects do not randomly identify a useful AI system at the beginning and then always stick with it. Second, we represent a large, dynamic market in which learning does not meaningfully alter the search set. Third, subjects learn about the market's lemon density through independent information signals across rounds. 
In our design, an AI that is a \enquote{lemon} delivers wrong suggestions 15\% of the time, while a \enquote{peach} delivers correct suggestions 90\% of the time. Consequently, the actual accuracy of the AI pool is bound to the lemon density: 70\% for low, 40\% for medium, and 10\% for high density. 
To control for potential ordering biases \cite{Nourani2021AnchoringBias}, we (i) applied block randomisation across the presentation of the three tasks, and (ii) generated 400 random permutations of the ten instances to ensure each participant encountered uniquely ordered task instances.

\subsection{\revision{\textbf{Implementation and User Interface Design}}}
\label{sec:interface_design}
\begin{figure*} [!t]
      \centering
    \includegraphics[width=\textwidth]{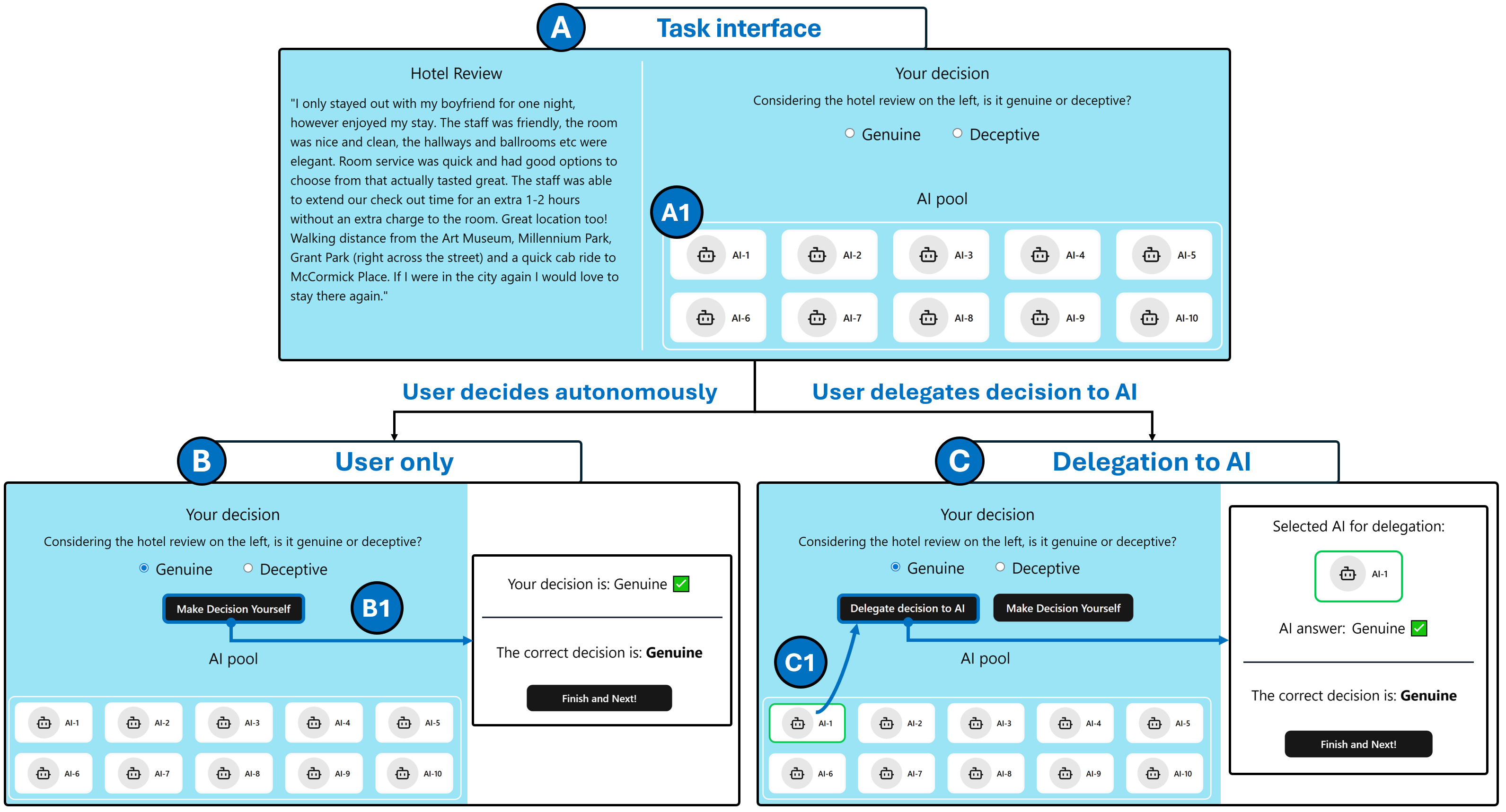}
  \caption{ \revision{Participants' flow for each trial in our user study, using a deceptive review detection task and no disclosure condition (A), with a pool of AIs available at the bottom right of the interface (A1). Participants had two options for decision-making: they could either complete the decision themselves or delegate it to one of the AIs. In the \enquote{User only} option (B), users made their own decision and received correctness feedback by pressing the corresponding button (B1). Alternatively, in the \enquote{Delegation to AI} option (C), users hovered over the AIs in the pool to reveal their accuracy and data quality, if available, based on the experimental condition. Then, participants selected one AI (highlighted in green) and could press the corresponding button to delegate the decision to the AI (C1), thereby receiving its prediction and feedback about its correctness.}
}
    \Description{
    Figure 2 illustrates the participant flow for each trial in the deceptive review detection task under the no disclosure condition. In panel A, participants view the main task interface, which includes a pool of AI systems at the bottom-right of the screen (A1). Participants choose between two decision-making options. In the \enquote{User only} option shown in panel B, participants make their own decision and receive correctness feedback after pressing the response button (B1). In the \enquote{Delegation to AI} option shown in panel C, participants can hover over AI tiles in the pool to reveal accuracy or data quality information when available, depending on the experimental condition. Participants then select an AI, which becomes highlighted in green, and press the delegation button (C1) to receive the AI’s prediction and feedback about its correctness.
    }
 
    \label{fig:interface_}
\end{figure*}

An example of the interface participants interacted with during the study is shown in Figure \ref{fig:interface_}. On each trial, users could hover over any AI to reveal the current state of information disclosure (see Fig. \ref{fig:interface_}-A1). This was a deliberate design choice to validate participants' engagement with the available information on AI system quality in the disclosure conditions. Additionally, they could either delegate the decision to an AI (Fig. \ref{fig:interface_}-B) or make the decision themselves (Fig. \ref{fig:interface_}-C). After each decision, whether made with AI or independently, participants receive trial-by-trial feedback on the correctness of their choice, allowing them to adjust their strategies through the tasks.
The online user interface was developed and hosted as a web application using the Next.js\footnote{http://next.js/} framework and deployed on Vercel\footnote{https://vercel.com/}. We used React\footnote{https://react.dev/} for the frontend implementation, tRCP\footnote{https://trpc.io/} for the backend, and Supabase\footnote{Supabase.com} for data storage with PostgreSQL.

\section{Study Design}
\label{sec:study_design}
To answer our research questions, we conducted a pre-registered between-subjects study\footnote{The pre-registration can be found here: 
\url{https://osf.io/95nbv/?view_only=031735e68ed04baeb54ba2c207fd7b13}.} comprising three \textit{lemon density} conditions (\lowden{}, \medden{}, and \highden{}) $\times$ two \textit{information disclosure} conditions (\nod{} and \partd{}), plus an additional condition with \highden{} and \fulld{}, which we used as a benchmark. We do not gather complete factorial data for the \fulld{} condition because, in theory, user behavior in this condition should be fully deterministic. Participants observe high-quality AI systems and always use them except if the participants believe themselves to be exceptionally good (i.e. $\geq 0.9$ accuracy). Thus, \fulld{} serves as a behavioral benchmark representing an ``optimal'' world in which information asymmetries can be institutionally solved. \revision{All data and code pertaining to the study are publicly available to ensure reproducibility.\footnote{Data and code for reproducibility: \url{https://osf.io/qsaun/overview?view_only=42c81874313d4fb9899786fac847d24e}}}

This section describes the variables and measurements collected during the study, as well as the recruitment policies for participants, statistical analysis setup, and study procedure.

\subsection{Variables}
\label{sec:variables}

\subsubsection{Independent Variables} 
\begin{itemize}
    \item \textbf{Lemon density} (\textit{categorical, between-subjects}). This variable controls the occurrences of AI \enquote{lemons} within the ten AIs pool for which participants can delegate the decisions to. We do not inform participants about the lemon density, although we do inform them that the share of AI systems remains constant throughout all tasks:

    \begin{itemize}
        \item \lowden{}: 3 out of 10 AIs are lemons (2 low accuracy, 1 high accuracy) = 70\% accuracy of AI pool.

        \item \medden{}: 6 out of 10 AIs are lemons (4 low accuracy, 2 high accuracy) = 40\% accuracy of AI pool.

        \item \highden{}: 9 out of 10 AIs are lemons (6 low accuracy, 3 high accuracy)  = 10\% accuracy of AI pool.
    \end{itemize}

    \item \textbf{Information disclosure} (\textit{categorical, between-subjects}). This variable controls how much information is disclosed for each AI system, considering both simulated accuracy on the test set and the data quality on which an AI is trained:

    \begin{itemize}
        \item  \nod{}: Participants only see AI names, as we do not disclose any information about each AI model.

        \item \partd{}: Participants only see the accuracy for each AI model.

        \item \fulld{}: Participants see the accuracy and data quality for each AI model.\footnote{Note that we exposed participants to the full disclosure condition only using high lemon density as a benchmark condition.}
    \end{itemize}

\end{itemize}

\subsubsection{Dependent Variables} 

\begin{itemize}
    \item \textbf{Delegation to AI} (\textit{continuous}). Percentage of participants' delegations to AI in the 30 trials.
    
    \item \textbf{Coins earned} (\textit{continuous}). The number of coins participants earned as a result of correct predictions across the 30 trials (i.e., proxy of task performance).
    
    \item \textbf{Delegation to Lemon AI} (\textit{continuous}). Percentage of participants' delegations to a \enquote{lemon AI} when they used the AI pool.



\end{itemize}

\subsubsection{Descriptive and Control Variables} 

\begin{itemize}

     \item \textbf{Task familiarity} (\textit{numerical, within-subjects}). We asked participants to state their familiarity with each task (i.e., loan prediction, deception detection, and skin cancer prediction) using a 5-point Likert scale from \enquote{1 - No experience} to \enquote{5 - Highly experienced}. 

    \item \textbf{Risk attitudes} (\textit{numerical}). We assessed participants' risk attitudes using Dohmen et al.'s ten-point scale \cite{Dohmen2011RiskAttitudes} by asking the following question: \textit{\enquote{In general, how willing are you to take risks?}} (1: \enquote{not at all willing to take risks}; 10: \enquote{very willing to take risks.}). 

    \item \textbf{Affinity for technology} (\textit{continuous}).
    This metric indicates the curiosity and willingness to engage with the technical workings of systems \cite{Langer2022ATIScale}. We measured it by taking the mean score of the four items (1: completely disagree, to 6: completely agree) in the Affinity for Technology Interaction (ATI) scale proposed by Franke et al. \cite{Franke2019AffinityforTechnologyInteractionScale}, consistent with prior work \cite{Langer2022ATIScale,Yurrita2023ATIScale,Yurrita2025AIliteracy}. 

    \item \textbf{AI literacy} (\textit{continuous}). 
    Because people have different motivations to engage with AI assistance \cite{Long2020AiLiteracy,ChunWeiMing2021AIliteracy,LEICHTMANN2023AILiteracy,FOROUDI2025AISensationAIliteracy,Yurrita2025AIliteracy}, we measured AI literacy by taking the mean of the four items that make up the self-assessed AI Literacy (AILIT) scale from Schoeffer et al. \cite{Schoeffer2022AiLiteracy}. Responses were collected on a 5-point Likert scale (1: strongly disagree; 5: strongly agree).\footnote{Self-Assessed AI Literacy (AILIT) full questions: \url{https://github.com/jakobschoeffer/facct22-130-appendix/blob/main/facct22-130-appendix.pdf}}

    \item \textbf{Perceived lemon density} (\textit{numerical}). We asked participants to respond about how many AIs were lemons in their opinion on an eleven-point numerical scale (0-10) after the fourth and ninth trials of each task.


\end{itemize}

\subsection{Participants}
\label{sec:participants}

\subsubsection{Recruiting and Filter Criteria}
We received ethics approval from the German Association for Experimental Economics Research e.V. and then proceeded to recruit participants from Prolific\footnote{https://www.prolific.com/\label{prolific_}} by following these selection criteria: equal participation across genders, age of 18 or more, high English proficiency, approval rate over 99\%, and Desktop as a mandatory participation device. 
We rewarded participants with \pounds 2.5 for completing the study, based on an average completion time of 25 minutes, which equates to an average of \pounds 6 per hour. We gave an extra \pounds 0.1 to participants for each correct response to encourage high-quality work. Altogether, participants were compensated an average of \pounds 4.4 for a 25-minute completion time, which corresponds to an hourly wage of \pounds 10.56, which is considered a fair payment on the Prolific platform~\cite{kaur2026incentive}. In total, we recruited 330 participants (50\% female), 50 for each \nod{} $\times$ Density and \partd{} $\times$ Density conditions, as well as 30 for our benchmark \fulld{} with \highden{} condition. We decided to recruit fewer participants in our benchmark condition because, theoretically, behavior in this condition is fully deterministic. Herein, participants with an expected performance of $< 0.9$ would always choose a high-quality AI system and delegate.

\subsection{Procedure}
\label{sec:procedure}
After obtaining informed consent, participants were assigned to one of the seven conditions based on the level of information disclosure and lemon density in a balanced fashion (except for the full disclosure, high lemon density condition, see Section \ref{sec:participants}).
Next, we collected participants' familiarity with each of the three tasks, their risk attitudes, as well as their affinity for technology and AI literacy, as assessed by questionnaires. During this phase, they also completed two attention checks and could further proceed with the study only if they obtained at least one correct answer.\footnote{We used valid attention checks allowed from Prolific where the answer was explicitly reported in the question text: (i) \enquote{For this question, please select the option "Largely disagree"}; (ii) \enquote{For this question, please select the option "Largely agree"}.}
Then, they completed a familiarization tutorial that explained the general purpose of the study and important terminologies such as AI's accuracy and data quality. The tutorial included two trials for each task: skin cancer prediction, loan approval, and deceptive hotel reviews, delivered with block randomization ordering, which was also be used for the next 30 trials.
The tutorial also outlined the option to choose an AI from a pool of ten to assist with current decisions, which could be accessed on demand by delegating the current decision to it. Therefore, after participants decide to select an AI for assistance during a trial, the pool of AI options will refresh for the next trial, while still respecting the current conditions related to information disclosure and lemon density (see Sec. \ref{sec:instance_selection}). 

Participants were then asked to answer some questions about the tutorial as comprehension checks, and only those who provided all the correct answers qualified to proceed with the user study. 
Afterwards, participants completed 30 trials, ten for each task, ensuring block randomization order across them. Specifically, upon entering a task, we provided participants with information about the dataset and the task's purpose, showing them one instance for the positive class and one for the negative class. For each decision, participants had the option to complete the trial themselves or delegate the trial's decision to an AI in the AI pool. Within each task, we additionally elicit subject beliefs about the market’s lemon density after rounds four and nine.
As a post-test, participants were informed about their total earned coins and asked for optional textual feedback about the study.

%% file: 4_Results.tex
\begin{figure*}[t]
    \centering
    \includegraphics[width=0.49\textwidth]{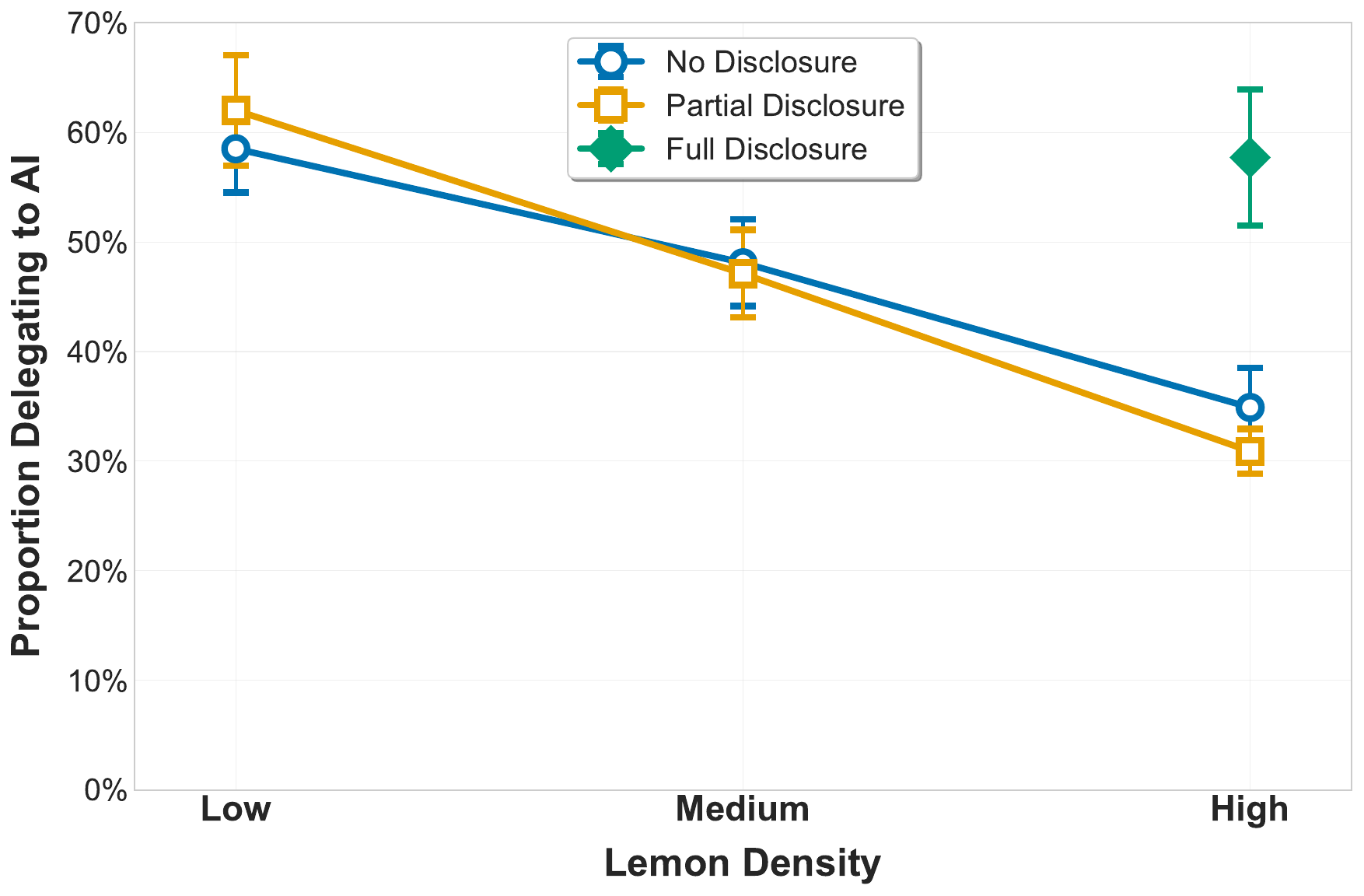}
    \includegraphics[width=0.49\textwidth]{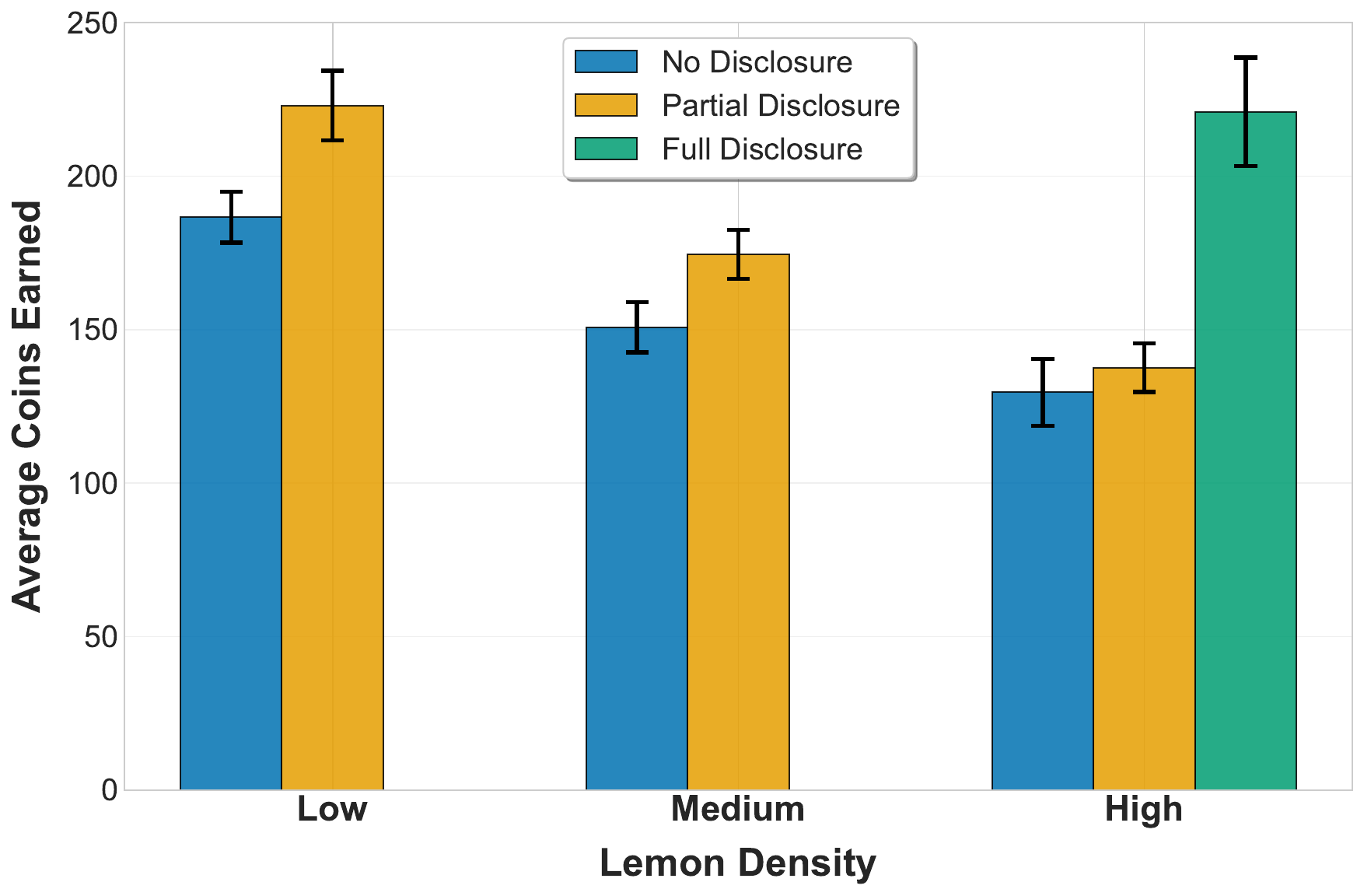}
    \caption{\textbf{Left:} Average delegation rates across the seven conditions collapsed \revision{at} task-level. \revision{There are no significant differences in delegation to AI between the \nod{} and \partd{} conditions across the three lemon densities, whilst \fulld{} only outcompetes \nod{} in terms of delegation rate in the \textit{High density} condition.} \textbf{Right:} Average coins earned per task across the seven conditions. \revision{Participants in \partd{} outperform those in \nod{} in terms of the number of coins earned, except for the \highden{} condition. Instead, the average coins earned in the \fulld{} condition is significantly higher than \nod{} and \partd{} conditions with a \highden{} of lemons.} Error bars represent 95\% confidence intervals.}

    \Description{
    Figure 3 presents two charts summarizing participants' delegation behavior and average coins earned across seven experimental conditions. The line chart on the left shows average delegation rates aggregated at the task level. Delegation to AI does not differ significantly between the No Disclosure and Partial Disclosure conditions across low, medium, and high lemon density. Full Disclosure shows higher delegation than No Disclosure only in the high-density condition. The bar chart on the right shows the average number of coins earned collapsed at the task level. Participants in Partial Disclosure earn more coins than those in No Disclosure in all conditions except the high-density condition. The average coins earned in the Full Disclosure condition are significantly higher than in the No Disclosure and Partial Disclosure conditions for high-density lemons.}
    \label{fig:r_1}
\end{figure*}

\section{Results and Analysis}

The final participants' sample comprised 330 users, with 166 females and 164 males, and an average age of $M$ = 37.44 and $SD$ = 12.55. Participants reported relatively low familiarity with all three tasks, and no differences were observed between the experimental conditions (see Table \ref{tab:task_familiarity} in the Appendix). Similarly, participants' risk propensity did not differ across conditions (Kruskal-Wallis: $\chi^2 = 2.88$, $df = 6$, $p = 0.82$).  
Further, participants reported an average Affinity for technology of $M$ = 3.88 ($SD$ = 0.58, 6-point Likert scale) and an average AI literacy of $M$ = 3.81 ($SD$ = 0.71, 5-point Likert scale).
\footnote{Shapiro-Wilk tests indicated significant deviation from normality for both measures (AI literacy: $W = 0.966$, $p < .0001$; affinity for technology: $W = 0.954$, $p < .0001$). Kruskal-Wallis tests showed no significant differences across conditions for AI literacy ($\chi^2(6) = 2.74$, $p = .84$), whereas affinity for technology differed across conditions ($\chi^2(6) = 13.4$, $p = .037$). Post-hoc pairwise comparisons (Bonferroni-adjusted) identified a single significant contrast for affinity for technology: \textit{full\_high} vs \textit{none\_medium} ($Z = 3.10$, $p = .0019$, $p_{\text{adj}} = .0403$).}

First, we look at the effect of information asymmetries without disclosure institutions on user AI adoption and performance. Figure \ref{fig:r_1} shows average delegation shares and performance (Coins earned) across the seven conditions. In \nod{}, delegation generally decreases with the density of lemons in the market. Over time (Figure \ref{fig:a1} in the Appendix), \revision{we found that participants} in \lowden{} increase delegation, but learning quickly stalls. In \medden{}, the trend is more ambiguous and even in \highden{}, the delegation shares remain comparably high. A mixed effect logistic panel regression (Table \ref{tab:a_01} in the Appendix) confirms only very limited treatment differences, where \revision{participants} learn to delegate more (less) in \lowden{} (\highden{}), but effect sizes are small. This is surprising, because \revision{participants} can learn over 30 rounds, and the differences between \lowden{} and \highden{} regarding expected value of AI delegation are quite large. Figure \ref{fig:beliefs} confirms that \revision{participant} beliefs $\lambda$ systematically deviate from true density values and do not significantly improve over time. In \lowden{}, \revision{participants} over-estimate the share of lemons, whereas they under-estimate them in \medden{} and \highden{}. Looking at \partd{} in Figure \ref{fig:r_1} reveals that overall, delegation rates are similarly imprecisely tuned to the actual distribution of low-quality AI systems. Despite similar delegation rates, the right panel shows that \revision{participants} in \partd{} substantially outperform those in \nod{} with regard to performance, \revision{providing support for hypothesis H4}. To quantify the effect of partial disclosure institutions on \revision{participants}' delegation behavior, we run a mixed effects logistic regression with participant-level random effects (Table \ref{tab:reg_1}), confirming no effect of \partd{}. However, in line with Bayesian information updating, partial disclosure does lead to significant and large efficiency increases (Table \ref{tab:reg_2}) in \lowden{} and \medden{}. Thus, disclosing accuracy information does not affect the prevalence of delegation, but the quality. Finally, in our benchmark condition  \fulld{} \highden{}, \revision{participants} only delegate 57.7\% of problems to the AI, despite being able to identify a high-quality system with certainty. This leads to substantial efficiency losses of 147 Coins on average.


\begin{figure}[t]
    \centering
    \includegraphics[width=.7\linewidth]{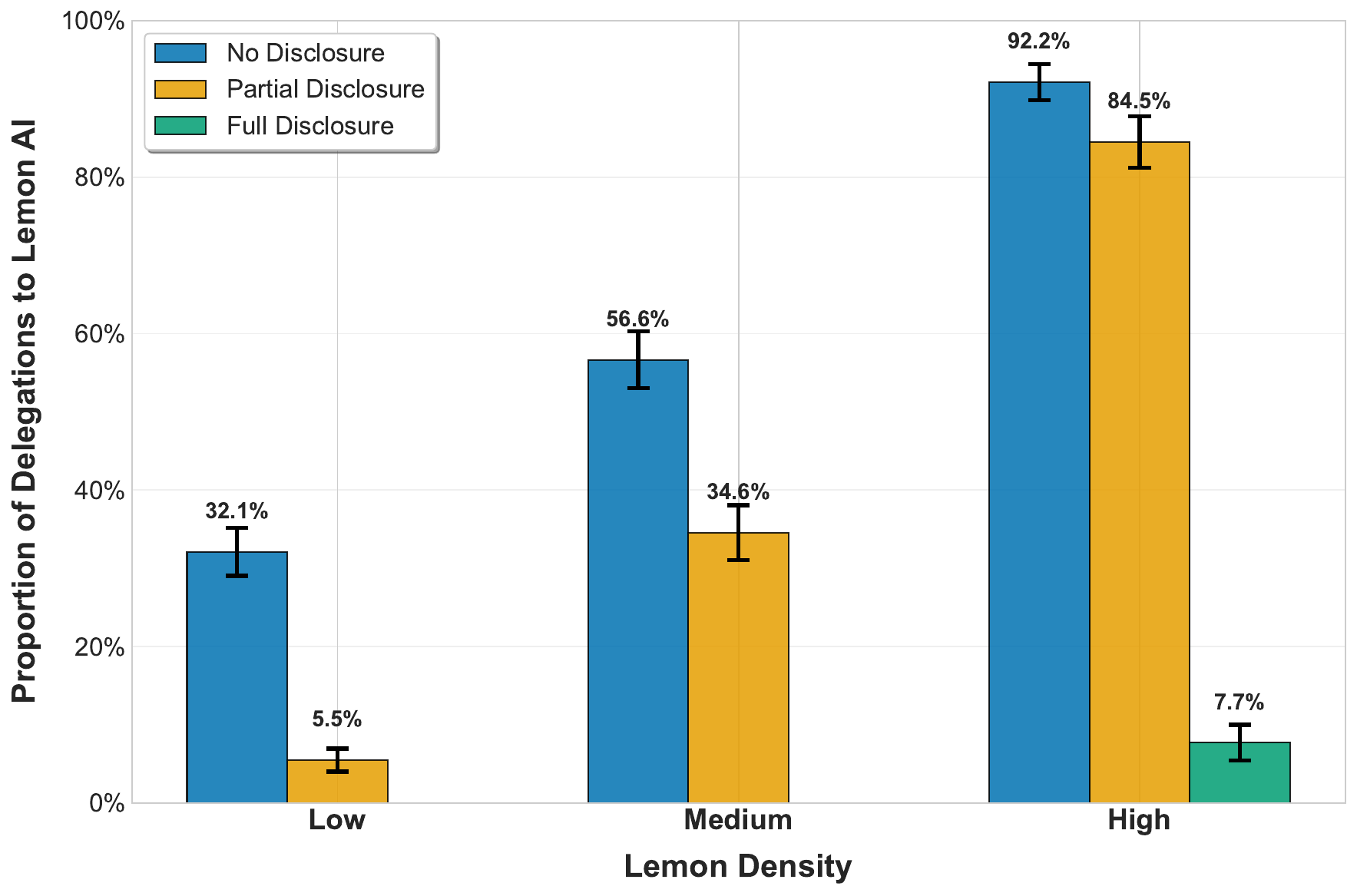}
    \caption{Share of delegation choices that targeted a lemon AI system. The bars only include choice data from observations in which a participant used the market \revision{(error bars represent 95\% confidence intervals). Participants in the \partd{} condition delegated significantly fewer decisions to lemon AIs than in the \nod{} condition across all lemon density conditions. Participants in the \fulld condition outperformed those in both \nod{}/\partd{} in terms of delegation rate in the \highden{} condition.} }
    \label{fig:r_2}
    \Description{
    Figure 4 shows a bar chart depicting the percentage of delegation choices directed toward a lemon AI system, collapsed at the task level. The bars include only trials in which participants chose to use the market. Participants in the Partial Disclosure condition delegated significantly fewer decisions to lemon AIs than those in the No Disclosure condition across all levels of lemon density. Full Disclosure resulted in even fewer delegations to lemon AIs and outperformed both No Disclosure and Partial Disclosure in the high-density condition.
    }
\end{figure}

\begin{table*}[t]
\centering
\caption{Mixed-effects logistic regression of delegation to AI (log-odds). Random intercept by participant. Task and time fixed effects included. Baseline disclosure is \emph{No Disclosure}.}
\label{tab:reg_1}
\begin{tabular}{lcccc}
\toprule
 & \textbf{Low density} & \textbf{Medium density} & \multicolumn{2}{c}{\textbf{High density}} \\
 & (No vs Partial) & (No vs Partial) & (No vs Partial) & (No vs Partial vs Full) \\
\midrule
\addlinespace[2pt]
\textbf{Disclosure (vs No)} & & & & \\
\quad Partial & 0.580 \;(0.495) & -0.113 \;(0.343) & -0.123 \;(0.222) & -0.103 \;(0.291) \\
\quad Full    & ---             & ---              & ---              &  1.288 \;(0.343)$^{***}$ \\
\addlinespace[4pt]
Task FE (domain) & Yes & Yes & Yes & Yes \\
Time FE (round)  & Yes & Yes & Yes & Yes \\
Observations & 3{,}000 & 3{,}000 & 3{,}000 & 3{,}900 \\
\bottomrule
\end{tabular}

\vspace{0.5em}
\footnotesize
\emph{Notes.} Coefficients are log-odds; standard errors in parentheses. Models are GLMMs with logit link, random intercept for participant ($1|\text{user\_id}$), task fixed effects (domain dummies), and time fixed effects (round dummies). Significance: $^{*}p{<}.05$, $^{**}p{<}.01$, $^{***}p{<}.001$. 
\end{table*}

\begin{itemize}
    \item [] \textbf{Result 1:} Under full information asymmetries, \revision{participants} exhibited limited learning about the prevalence of low-quality AI systems. This leads to user adjustments in AI adoption that are directionally correct, but substantially too small.
\end{itemize}

\begin{itemize}
    \item [] \textbf{Result 2:} Partial disclosure institutions increase the efficiency of Human-AI collaboration under information asymmetries as long as the density of lemons is not too high.
\end{itemize}

To test the proposed mechanism of our model, Figure \ref{fig:r_2} illustrates the share of delegation choices in which participants selected a lemon AI. For \nod{}, the percentage is close to the actual distribution of lemons. Participants in \partd{}, on the other hand, are much more likely to avoid the low-quality systems across \textit{all} disclosure conditions. A random effects panel logit regression (Table \ref{tab:reg_3}) confirms that the partially informative accuracy labels allow participants to be significantly more likely to avoid AI lemons. People are generally able to improve their decision-making even when they only receive incomplete disclosure information. However, this effect appears to decrease in the density of low-quality AIs. While lemon shares are basically optimal in \lowden{}, they are 15 percentage points ``too high'' in \medden{} and 55 percentage points ``too high'' in \highden{}. This could be due to selection effects, where more engaged or capable \revision{participants} quickly learn to never delegate in \highden{}, or cognitive effort, which plausibly increases with the share of ambiguous accuracy label signals and thus potentially inhibits the efficiency of the Bayesian updating process in the context of information asymmetries.

\begin{itemize}
    \item [] \textbf{Result 3:} In line with the predictions of Bayesian updating under information asymmetries, participants who receive partial disclosure information are significantly less likely to rely on a lemon AI.
\end{itemize}

\begin{itemize}
    \item [] \textbf{Result 4:} The positive effect of partial information disclosure on AI system selection decreases with the density of low-quality AI systems.
\end{itemize}

\begin{table*}[t]
\centering
\caption{OLS (user-clustered) regressions of task coins by disclosure and density. Task fixed effects included; baseline disclosure is \emph{No Disclosure}.}
\label{tab:reg_2}
\begin{tabular}{lcccc}
\toprule
 & \textbf{Low density} & \textbf{Medium density} & \multicolumn{2}{c}{\textbf{High density}} \\
 & (No vs Partial) & (No vs Partial) & (No vs Partial) & (No vs Partial vs Full) \\
\midrule
\addlinespace[2pt]
\textbf{Disclosure (vs No)} & & & & \\
\quad Partial & $36.400^{***}$\,(7.159) & $23.800^{***}$\,(5.835) & 8.000\,(6.878) & 8.000\,(6.871) \\
\quad Full    & ---                      & ---                      & ---            & $91.400^{***}$\,(10.531) \\
\addlinespace[4pt]
Task FE (domain) & Yes & Yes & Yes & Yes \\
SE type & Clustered (user) & Clustered (user) & Clustered (user) & Clustered (user) \\
\addlinespace[2pt]
Observations & 300 & 300 & 300 & 390 \\
\bottomrule
\end{tabular}

\vspace{0.5em}
\footnotesize
\emph{Notes.} Coefficients; cluster-robust standard errors in parentheses. Baseline disclosure is \emph{No}. 
Models include task fixed effects (domain dummies). Significance: $^{*}p{<}.05$, $^{**}p{<}.01$, $^{***}p{<}.001$.
\end{table*}

\begin{table*}[t]
\centering
\caption{Mixed-effects logistic regression of selecting a lemon among delegations (log-odds). Random intercept by participant. Task and time fixed effects included. Baseline disclosure is \emph{No Disclosure}.}
\label{tab:reg_3}
\begin{tabular}{lcccc}
\toprule
 & \textbf{Low density} & \textbf{Medium density} & \multicolumn{2}{c}{\textbf{High density}} \\
 & (No vs Partial) & (No vs Partial) & (No vs Partial) & (No vs Partial vs Full) \\
\midrule
\addlinespace[2pt]
\textbf{Disclosure (vs No)} & & & & \\
\quad Partial & $-2.377^{***}$\,(0.235) & $-0.926^{***}$\,(0.115) & $-0.803^{***}$\,(0.222) & $-0.868^{*}$\,(0.362) \\
\quad Full    & ---                     & ---                     & ---                     & $-5.807^{***}$\,(0.464) \\
\addlinespace[4pt]
Task FE (domain) & Yes & Yes & Yes & Yes \\
Time FE (round)  & Yes & Yes & Yes & Yes \\
Observations & 1{,}808 & 1{,}428 & 988 & 1{,}507 \\
\bottomrule
\end{tabular}

\vspace{0.5em}
\footnotesize
\emph{Notes.} Coefficients are log-odds; standard errors in parentheses. Models: GLMM (logit) with random intercept for participant ($1|\text{user\_id}$), task fixed effects (domain), and time fixed effects (round). Significance: $^{*}p{<}.05$, $^{**}p{<}.01$, $^{***}p{<}.001$.
\end{table*}

Now, we focus on the rationality of participant choices, assuming risk neutrality. We are primarily interested in whether partial disclosure causes diverging delegation trends between \revision{participants} who outperform and those who under-perform the AI market (i.e. expected-payoff maximizing behavior), or if performance improvements are due to more targeted delegation choices independent from the user's accuracy. Because participants may have different abilities across tasks, we run task-separate mixed-effect logit panel regressions. Our main target variable interacts the disclosure condition with a dummy variable capturing whether a participant's expected payoff based on their own performance exceeds the expected payoff of entering the market (and making an informed choice in \partd{} and \fulld{}). All tables are shown in the Appendix (Tables \ref{tab:a1} -- \ref{tab:a3}). The interaction term does not have explanatory value for any of the three tasks, suggesting that partial disclosure information does not increase the rationality of users. Instead and in line with the delegation behavior plotted in Figure \ref{fig:r_1}, both the delegation shares and the delegation distribution across skill levels is largely equivalent between \nod{} and \partd{}. Hence, performance improvements are driven by \textit{more targeted} delegation choices, where participants use the accuracy label to successfully avoid lemon products. Beyond that, Tables \ref{tab:a1} -- \ref{tab:a3} also confirm that \revision{participants} with more accurate predictions than the AI market are generally less likely to use an AI system for deceptive hotel reviews and loan prediction. Thus, participants are sensitive to their own relative performance level during delegation choices. Regarding ``blind'' utilization of the market where participants always delegate, an OLS regression with clustered standard errors on the participant level (Table \ref{tab:a4}) finds very limited evidence for treatment differences. Mostly and in line with expectations, \fulld{} leads to a significant increase of full delegators. This makes sense, because participants can perfectly identify a high-quality model. \partd{} shows no consistent effect, but may increase full market usage in \lowden{}, i.e. when the density of high-quality AIs is high.

\begin{itemize}
    \item [] \textbf{Result 5:} Partial disclosure institutions do not affect the share of participants who maximize their expected payoff.
\end{itemize}

\begin{itemize}
    \item [] \textbf{Result 6:} Partial disclosure institutions increase average user performance and market efficiency by allowing delegators to make more informed delegation choices. 
\end{itemize}

Finally, we analyze how information asymmetries and participant-level characteristics affect AI use and Human-AI collaboration efficiency more generally. To that end, we deploy a full mixed effect logistic panel regression with all conditions and our main co-variates technology affinity, risk attitudes, AI literacy, and task familiarity. Results (Table \ref{tab:a5}) confirm that delegation decreases with lemon density, \revision{thus supporting hypothesis H2}, but increases over the course of each task. Participants adapt directionally correct, albeit too little, and learn over time. In addition, we find significant negative correlations between task familiarity and delegation, as well as affinity for technology and delegation, while risk attitudes and AI literacy do not explain behavior. Looking at performance (Table \ref{tab:a6}), in contrast, suggests that participants with more risk-loving preferences tend to make fewer correct predictions, while AI literacy correlates positively with income. Panel 2 adds performance and delegation variables, which illustrate that people tend to over-rely on AI in \nod{} and thereby reduce their potential income. For \partd{}, on the other hand, \revision{participants} under-rely on AI in \lowden{}, and over-rely in \highden{}. Notably, increasing the lemon density to medium does not negatively affect user performance under the partial information disclosure condition, underscoring the usefulness of the partial signal, as it almost entirely offsets the increase in low-quality systems. With high density, people always strongly over-delegate in the absence of full disclosure. Overall, participant choices are only imperfectly calibrated to the decision environment under information asymmetries. They delegate too little if lemons are rare, and delegate too much when lemons are common. \revision{Participants} efficiently exploit partial information signals, causing more targeted delegation choices. However, rationality does not improve. Fully disclosing all relevant variables has a strong and positive effect on efficiency, but is severely constrained by \revision{participants}' tendency to rely on themselves.


%% file: 5_Discussion.tex
\section{Discussion}
This paper experimentally analyzes the effect of information asymmetries in the presence of low-quality AI systems (\textit{lemons}), on user behavior, and the concurrent efficacy of (imperfect) disclosure. It is notoriously difficult to solve informational market frictions through regulatory interventions, as full transparency is often neither an achievable, nor desirable goal. The market for AI systems illustrates this problem very well. Regulations differ substantially between countries, leave a lot of room for interpretation, and are subject to constant debates between politicians, regulators, providers and citizens. In addition, there has so far been very little empirical work about the behavioral validity of different interventions, such as (partial) disclosure rules (see, e.g., EU AI Act), which adds to the general confusion of the debate. Specifically in the context of disclosure, one main driver behind the efficacy of forcing or encouraging providers to reveal relevant information is the behavior of users. In order to be useful, people must be able to integrate the relevant information signals into their choice process. This is, arguably, a lot easier when users have full information about all AI systems, because it enables simple cost-benefit comparisons. In reality, however, people must handle a lot more ambiguity, and follow imprecise signals that may affect optimal choices, but are not sufficiently complete to render behavior deterministic. Hence, improvements are subject to users' belief updating processes and, by extension, rationality. This tension is the main focus of our paper.  

\vspace{.5em}
\noindent
\begin{quote}
\textit{RQ1: How do information asymmetries about AI system capabilities affect \revision{users' adoption of AI} and market outcomes in the presence of low-quality AI systems?}    
\end{quote}

\revision{Information asymmetries are detrimental to 
the outcomes and the overall efficiency of Human-AI collaboration. Participants in our study over-reacted to uncertainty when the density of lemons was low, and under-reacted when the density was high. The quality of delegation choices improved at the beginning, but quickly stalled, suggesting limited learning under uncertainty. In line with this, participants' beliefs about the market's lemon density were stable over time (Figure \ref{fig:beliefs}). Under- and over-utilization of the AI systems was common. As a result, utilization of low-quality AI systems remained high, and efficiency is mostly a function of lemon density.}
\revision{A central finding of this work is the persistent under-reliance on AI systems even in the \fulld{} condition, where participants could identify a high-quality AI system with certainty. This suggests that transparency alone is insufficient to guarantee optimal adoption. Several psychological and cognitive factors may explain this behavior. First, overconfidence in self-judgment can lead users to favor their own reasoning over algorithmic recommendations~\cite{He2023DunningKruger}, even when objective evidence indicates superior system performance. Second, users may exhibit a strong preference for agency and control, consistent with prior HCI research showing that autonomy is often prioritized over efficiency in decision-making contexts~\cite{mehrotra2024systematic}. Third, residual distrust of AI systems, stemming from concerns about hidden risks or accountability, may persist even when technical information is fully disclosed, reflecting challenges in trust calibration.}

\revision{These explanations align with established frameworks such as trust in automation \cite{lee2004trust,mehrotra2024systematic}, which emphasize that trust is shaped by more than objective reliability, and bounded rationality \cite{kaur2024interpretability}, which highlights cognitive effort and satisficing strategies under uncertainty~\cite{salimzadeh2024dealing}. From an HCI perspective, this finding underscores the need for interaction designs that go beyond transparency to actively support trust calibration and reduce cognitive burden. }

\vspace{.5em}
\noindent
\begin{quote}
\textit{RQ2: How do different information disclosure requirements about AI system capabilities impact \revision{\textit{user behavior}}, market outcomes, and reliance on \revision{\textit{low-quality AI systems}}?}    
\end{quote}

We document four main results. One, partial disclosure substantially increases the efficiency of human-AI collaboration in the presence of \revision{low-quality AI systems}. Two, this effect is driven by more effective delegation through the marginalization of \revision{low-quality AI systems}, rather than increased rationality. Over- and under-delegation rates remain constant, but delegation overall improves. Three, high density \revision{of low-quality AI systems} alleviates the positive effect of partial disclosure, as participants are significantly less effective in utilizing the provided information. Fourth, even under a full disclosure institution that entirely eliminates market information asymmetries, \revision{participants} exhibit strong AI under-reliance, leading to large efficiency losses \revision{in decision-making}.

\noindent\revision{\textbf{AI qualities as a vehicle for information disclosure.}}
\revision{In this work, we leverage AI accuracy as a cue for partial disclosure and incorporate additional data quality indicators to enable full disclosure. 
These choices reflect the need for interpretable, actionable signals that help lay users make informed decisions during interactions.
Our approach is generalizable and grounded in human-centered AI literature on decision-making, where model-centric cues (e.g., AI accuracy, confidence, and calibration) are widely recognized for their influence on user reliance attitudes \cite{Yin2019AIAccuracyOnTrust,Zhang2020ConfidenceExplanationsAccuracyTrust,Rechkemmer2022AIPerformanceAndConfidenceEffectsOnUsers,he2023stated,Kahr24AccuracyHumanExplanations,Cao2024UncertaintyPresentation,Cau2025CuriosityTraits}. 
In contrast, \enquote{data-centric explanations}, which foreground properties of the training data, have only recently emerged in XAI literature.
These explanations, when combined with model-centric ones, can form hybrid strategies that support richer, more transparent interactions, helping foster appropriate reliance on AI systems \citep{Anik2021DataCentric,zha2025datacentricartificialintelligencesurvey,Bhattacharya2023DirectiveExplanations,Bhattacharya2024EXMOS,Cau2025CuriosityTraits}. 
Future work should investigate how different combinations of partial- and full-disclosure cues shape user perceptions, trust calibration, and delegation behaviors across diverse contexts, tasks, and stakes. Such studies will be critical for designing disclosure mechanisms that reduce information asymmetry and support equitable, user-centered decision-making in real-world interactive systems.}

\subsection{Implications}

 Our work bears important implications for the future design and regulation of AI systems. 
 \revision{As AI systems proliferate across consumer and professional contexts, our results underscore the critical role of information disclosure in shaping user reliance behaviors. Our insights demonstrate that information asymmetries between system providers and end-users do not merely create market inefficiencies, but they also lead to miscalibrated trust, poor delegation strategies, and suboptimal task outcomes.  Users in our study rarely learned enough to identify environments with high versus low prevalence of low-quality systems, resulting in persistent patterns of over-reliance or under-reliance. This corroborates recent work in the HCI community that has argued that misaligned mental models can undermine trust and decision-making, and that grounding interactions in shared understanding is essential~\cite{amershi2019guidelines,bansal2019beyond,hoffman2023measures,andrews2023role}.}

\revision{Our findings suggest that even partial disclosures can meaningfully improve user decision-making, provided they are interpretable and actionable. HCI researchers and practitioners must prioritize mechanisms that surface AI accuracy, uncertainty, provenance, and limitations, even if the information eventually disclosed to users of such systems is incomplete. From a regulatory and policy perspective, prioritizing enforceable disclosure rules, even if incomplete, can be important. Policy and regulatory efforts could aim to enforce minimum disclosure standards for AI systems that mandate interpretable cues (e.g., accuracy and data provenance) rather than exhaustive technical details, which can also be difficult to enforce or audit.
Preventing selective reporting or ``gaming'' of transparency by AI system developers, providers, or suppliers would require enforceable strategy-proof formats (e.g., standardized templates, or machine-readable labels).} 

\revision{From the perspective of design implications for human-AI interaction, our findings underline the importance of embedding meaningful information disclosure cues near decision points (e.g., alongside AI advice or assistance) rather than in separate documentation or other pathways. For example, providing short tooltips or expandable panels that explain to users what accuracy and data quality mean for the task at hand can help surface the required disclosure signals in situ. Filtering mechanisms can enable users to select AI systems based on quality tiers, supporting user agency and control. In addition, incorporating dynamic feedback mechanisms that respond to user behavior (e.g., nudging users when persistent under-reliance is detected, or highlighting missed opportunities for efficiency gains) can help support trust calibration without forcing delegation. }

\revision{Collectively, these results reinforce the HCI imperative to design for informed interaction~\cite{shneiderman2022human,stephanidis2025seven,bach2024systematic,cavalcante2023meaningful}, where users can meaningfully interpret and act upon AI outputs, even in the presence of systemic opacity. Rather than aiming for exhaustive transparency, designers and regulators should prioritize strategically minimal disclosures that help users form accurate mental models.}


\subsection{\revision{\textbf{Caveats, Limitations, and Future Work}}}
\label{sec:limitations_future_work}

\textbf{Static Market Dynamics.} In this study, we rely on a market without dynamic exit and entrance of sellers and consumers, as well as fixed seller behavior. \revision{By holding market dynamics constant, we can focus on how users interpret and act on the disclosure signals in a controlled environment. This provides actionable insights for designing trust cues, information labels, and decision-support interfaces that help users make informed choices under uncertainty. However, it is worth noting that several real-world AI marketplaces are dynamic, in which users’ choices can influence sellers' behavior and vice versa.} While our setup allows us to focus on consumer behavior, long-term dynamics, price-setting and market equilibria are beyond the scope of this paper. Future research may expand on our results and framework to gather more nuanced and generalizable results through market experiments. \revision{Future work should also explore interactive and adaptive disclosure designs that remain effective when market conditions change, ensuring transparency and appropriate adoption of AI systems at scale.}

\noindent
\textbf{Operationalization of AI Quality.} 
Our study relies on simulated AIs with relatively fixed accuracies with only two relevant quality dimensions (accuracy on the test set and data quality in the training set).  While performance cues and training data quality are powerful and simple signals for lay users when considering AI models (e.g., similar to model cards \revision{or AI nutrition labels} \citep{bender2018data,mitchell2019model,gebru2021datasheets,shen2022model,kennedymayo2025reclassifying,stoyanovich2019nutritional}), future work may expand on our results by deploying ``real'' AI systems and introducing higher-dimensional quality attributes. This, in particular, may interact with how people process information signals, as they directly affect the complexity and kind of information users observe. 
\revision{However, increasing AI quality cues should account for both lay and expert users. Presenting too many AI qualities may overwhelm lay users and increase overconfidence biases such as inflated self-assessment (e.g., the Dunning-Kruger effect \cite{He2023DunningKruger}), leading to misunderstandings and inappropriate reliance on AI. Expert users, instead, can benefit from richer information, including data inspection (e.g., intrinsic data biases, data quality, and processing) \citep{Anik2021DataCentric,Bhattacharya2024EXMOS,Cau2025CuriosityTraits} and AI indicators (e.g., bias propagated to the model, fairness principles, and confidence calibration) \cite{SilvaFilho2023Calibration,Wang2023Bias,Cao2024UncertaintyPresentation,Ma24UserConfidenceCalibration}. In addition, further studies should test AI quality disclosures across different task stakes and difficulties to understand how various indicators influence real-world delegation and market outcomes.}


\noindent
\revision{\textbf{On the nature of `lemons.'} While our study operationalizes low-quality AI systems (i.e., lemons) through accuracy and data quality, real-world AI systems exhibit additional quality dimensions such as fairness, robustness, and safety. These characteristics are often harder to quantify and communicate because they depend on context, edge cases, or adversarial conditions. If lemons were defined by these properties, the dynamics of user interaction with the AI systems and their reliance on AI advice might shift significantly. Users could underestimate risks even under full disclosure, or struggle to interpret complex ethical trade-offs. This would imply that disclosure mechanisms must evolve beyond numeric performance indicators to include interpretable signals for fairness and safety; potentially through scenario-based explanations, visual risk cues, or tiered disclosure interfaces. This is an important avenue for future research. Broadening the scope of quality dimensions is essential for designing transparency strategies that support informed human-AI interaction in ethically sensitive domains.}

\noindent
\revision{\textbf{Task complexity.} In our study, the decision-making tasks were cognitively manageable, which may have allowed participants to devote more attention to evaluating the information disclosed about the AI systems. In more complex or constrained contexts (e.g., under time pressure)~\cite{salimzadeh2023missing,salimzadeh2024doubt,cao2023time,swaroop2024accuracy}, users may have fewer cognitive resources available for scrutinizing AI quality signals, which could amplify the negative effects of information asymmetry and even diminish the benefits of full disclosure.}

\noindent
\textbf{Delegation.} 
\revision{We made a deliberate choice to model delegation as a binary decision (i.e., relying on AI or not). This offers a clear, interpretable starting point for studying reliance behaviors under controlled conditions. This simplification is common in human-centered AI research~\cite{erlei2024understanding,biswasEG25} because it allows us to isolate the core phenomenon (information asymmetry and disclosure effects) without introducing confounding factors from more complex interaction patterns. We can thereby systematically test how disclosure strategies and lemon density influence adoption decisions, which would be difficult to disentangle in a continuous reliance model.
Moreover, binary delegation reflects real-world decision points in many contexts, such as whether to accept an AI recommendation or proceed manually (e.g., approving a credit score, accepting a medical triage suggestion). While actual workflows often include verification or partial reliance, these behaviors are layered on top of the fundamental choice to delegate or not. Understanding this baseline is essential before designing for more nuanced behaviors. Having said that, future research should explore experimental designs that allow for graded reliance (e.g., sliders for confidence weighting), verification actions (e.g., optional checks before committing), and override mechanisms to represent other real-world workflows.}

\revision{Furthermore, additional metrics focusing on different users' perceptions and traits could provide a more nuanced understanding of our results. For example, in light of participants' under-delegation in the full disclosure high-density condition, future research could incorporate measures of individual characteristic that are relevant for explaining delegation behavior, such as human decision confidence \cite{Miller2015OVerconfidenceUnderconfidence,Pescetelli2021DecisionConfidenceTrust,FosterAndRenie2024HumanConfidenceCalibrationStudents,Cau2025CuriosityTraits,li2025confidence}, perceived trust and autonomy \cite{Ma2023CorrectnessLikelihoodAIUsersIncome,Ma24UserConfidenceCalibration,He2025ConversationalXAIOnDemand}, or algorithm aversion (i.e., the tendency to reject the suggestions made by algorithms, even when those algorithms outperform human judgment) \cite{Dietvorst2014_alg_aversion,Germann2023_alg_aversion,jin2025_alg_aversion}.} 
This, along with people's actual ability to appropriately rely on AI systems, may induce profound changes in the effectiveness of disclosure regimes.

\noindent
\textbf{Costless Verification.} Real world disclosure regimes are noisy, costly, and exhibit variance in credibility and trust. We abstract from these factors, but acknowledge that these factors will play a role for actual disclosure regulations. This includes sellers' strategic verification choices, price-setting, endogenous seller competition via consumer choices, and, in equilibrium, what kind of disclosure level emerges from strategic market interactions. It is, for instance, not guaranteed that in equilibrium, people can fully trust partially informative information labels (in our case, we select a publicly known and fixed probability of a wrong signal, which reveals the usefulness of the label fully).

\revision{Future research should investigate why users under-delegate even under full disclosure. Qualitative or mixed-methods studies could unpack underlying factors such as trust calibration, perceived control, and accountability concerns, offering richer insights into cognitive and social dynamics that shape reliance on AI~\cite{gadiraju2025enterprising}. Building on these findings, we aim to translate behavioral patterns into actionable interaction designs in the near future. For example, tiered disclosure interfaces that progressively reveal information~\cite{springer2019progressive}, or adaptive trust cues that respond to user behavior in real time. Expanding our operationalization of AI quality beyond accuracy and data generalizability to include fairness, robustness, and safety will further align disclosure strategies with emerging ethical and technical standards~\cite{el2024transparent,laux2024three}.} 

%% file: 6_Conclusions.tex
\section{Conclusion}
\label{sec:conclusion}
This paper experimentally demonstrates that information asymmetries fundamentally undermine efficient AI adoption in consumer markets. When faced with uncertainty about AI quality across a pool of opaque systems, participants adjust reliance in the right direction over time, but still under-use AI when lemons are rare and over-use AI when lemons are common. Partial disclosure markedly improves efficiency by steering delegators away from lemons, even though it does not increase the share of “rational” adopters. Under full disclosure, users still under-delegate, leaving sizable performance on the table. Together, these findings suggest that enforceable, even imperfect, disclosure rules can deliver meaningful welfare gains in real-world AI markets that suffer from information asymmetries. They underscore the critical role of institutional design in shaping AI adoption patterns and highlight the potential for targeted regulatory interventions to mitigate market failures in the growing AI economy. \revision{Our work introduces a novel experimental framework that adapts the classic \enquote{market for lemons} theory to study AI adoption under information asymmetry. By operationalizing lemon density as a design variable, we systematically examine how disclosure strategies interact with uncertainty to shape user reliance. This approach not only bridges economic theory and HCI but also offers an extensible paradigm for future research on trust calibration, transparency, and the design of human interaction with AI systems.}